\documentclass[superscriptaddress, twocolumn, prl, aps, longbibliography, nofootinbib]{revtex4-2}

\usepackage{amsmath, amssymb, color, wasysym, esint}
\usepackage{graphicx}
\usepackage{bm}
\usepackage{xcolor}
\usepackage{xspace}
\usepackage{float}
\usepackage{lineno}

\definecolor{linkcolor}{rgb}{0,0,0.6} 
\usepackage[pdftex,colorlinks=true,
	pdfstartview = FitV,
	linkcolor    = linkcolor,
	citecolor    = linkcolor,
	urlcolor     = linkcolor,
	hyperindex   = true,
	hyperfigures = false]{hyperref}

\begin{document}
\newcommand{\sinc}{\text{sinc}}

\title{Topology of pulsating active matter: Defect asymmetry controls emergent motility}

\author{Luca Casagrande}
\email{luca.casagrande@uni.lu}
\affiliation{Department of Physics and Materials Science, University of Luxembourg, L-1511 Luxembourg City, Luxembourg}

\author{Alessandro Manacorda}
\affiliation{Department of Physics and Materials Science, University of Luxembourg, L-1511 Luxembourg City, Luxembourg}
\affiliation{CNR Institute of Complex Systems, Uos Sapienza, Piazzale A. Moro 5, 00185 Rome, Italy}

\author{\'Etienne Fodor}
\email{etienne.fodor@uni.lu}
\affiliation{Department of Physics and Materials Science, University of Luxembourg, L-1511 Luxembourg City, Luxembourg}

\begin{abstract}
In pulsating active matter, topological defects are motile despite the absence of any macroscopic flows and microscopic self-propulsion. We reveal that this motility arises from a ratchet effect: the mechanochemical coupling between local oscillations and repulsive interactions breaks both spatial and time-reversal symmetries, thus leading asymmetric rotating defects to drift under fluctuations. This mechanism regulates a crossover between spiral waves connecting slow defects and fiber-like waves connecting fast defects, in analogy with the onset of heart rhythm disorder in cardiac tissues. We rationalize this crossover in terms of a fluctuating hydrodynamics that captures how motile defects spontaneously nucleate and move within an ordered background.
\end{abstract}

\maketitle



\begin{figure}[b] 
	\centering 
	\includegraphics[width=.8\linewidth, trim=0 0 0 0cm, clip=true]{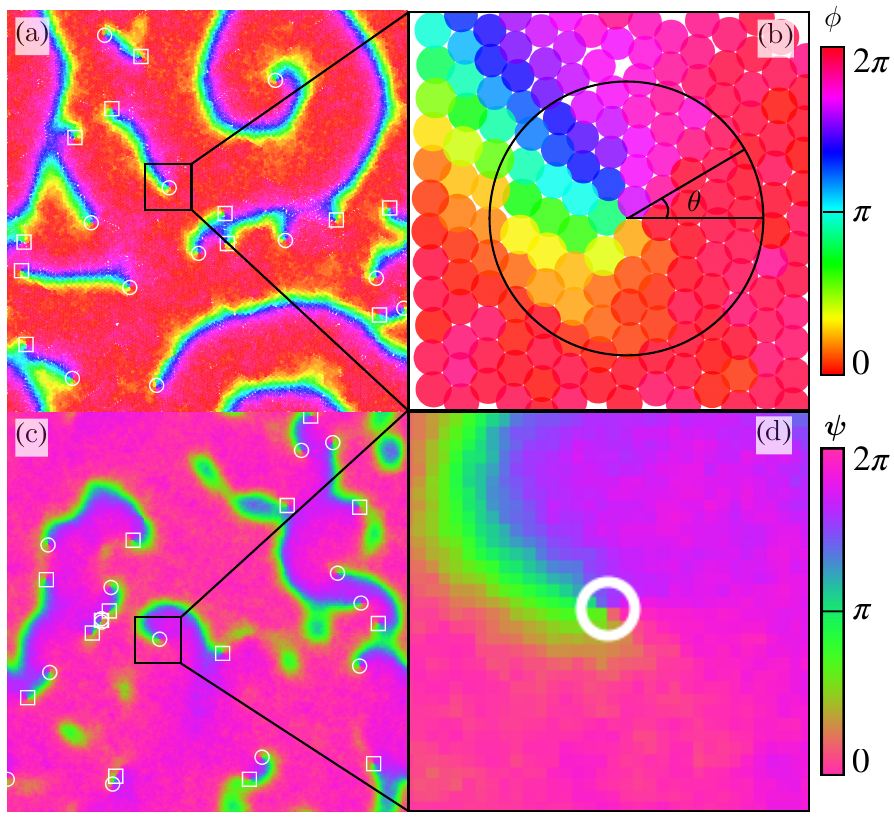} 
	\caption{Deformation waves connect defects that rotate with asymmetric cores for (a,b)~pulsating particles [Eq.~\eqref{eq:pam_part}], and (c,d)~fluctuating hydrodynamics [Eq.~\eqref{eq:hydro}]. The rotation is set by defect charges $+1$ ($\circ$, CW) and $-1$ ($\square$, CCW), while profile asymmetry is similar for $\pm 1$ defects. Parameters in Ref.~\cite{sm}.}
\label{fig1} 
\end{figure}

\begin{figure} 
  \centering 
  \includegraphics[width=\linewidth, trim=0 0cm 0 0, clip=true]{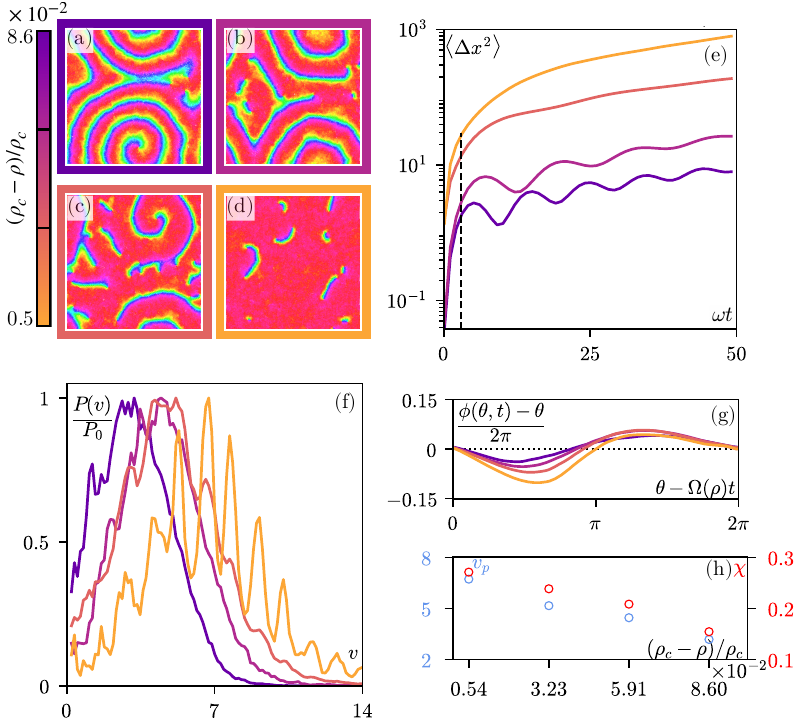}
  \caption{Defect statistics for pulsating particles [Eq.~\eqref{eq:pam_part}]. (a)--(d)~Phase patterns change from spirals to fiber-like waves as density $\rho$ increases; color code and parameters as in Fig.~\ref{fig1}(a).
  (e)~Mean-squared displacement $\langle\Delta x^2\rangle$ of defect cores as a function of time $t$; colors refer to $\rho$ values, and dashed line to time interval at which velocity is estimated.
  (f)~Defect velocity distribution $P(v)$ of  scaled by $P_0 = \max P(v)$.
  (g)~Defect profile $\phi(\theta,t)$ in the co-rotating frame $\theta-\Omega(\rho)t$.
  (h)~Peak velocity $v_p = {\rm argmax} P(v)$ and asymmetry factor $\chi[\phi]$ [Eq.~\eqref{eq:chi}].}
\label{fig2} 
\end{figure}

\textit{Introduction}.---Pulsating active particles subject to periodic mechanical deformations have garnered increasing attention~\cite{Tjhung2017, Koyano2019, togashi2019, parisi2023, li2024fluidization, zhang2023, liu2024, pineros2024, tirthankar2024, tirthankar2025, goth2025, zhu2025, Zhuli2025}. They capture collective effects in deformation-driven soft media~\cite{manning2023}, some of which play an important role in living systems~\cite{chiou2016, ishihara2017, Ikeda2019, Staddon-PlosOne2022, boocock2023, Shiladitya2024, tang2025, koehler2026}. For instance, excitation waves in cardiac tissues organize dynamical patterns around topological defects~\cite{karma2013, karma2022, Rappel2022}. During cardiac arrhythmia, the crossover from tachycardia to fibrillation changes these patterns from stable spirals with steady defects to fiber-like waves with motile defects~\cite{kim2007}. Similarly, deformation waves in pulsating active matter yield motile defects connected by spiral and fiber-like waves [Figs.~\ref{fig1}-\ref{fig2}]. The principles underlying these defect dynamics have remained largely elusive.

Defects help rationalize the mechanisms of pattern formation~\cite{bray1994, shankar2022, tubiana2024, Tang2025_b}. In cardiac and excitable systems, defect dynamics has been characterized theoretically and experimentally~\cite{barkley1995, garcia1999, biktashev2003, cameron2012, pravdin2023, roth2001, grill1995, mikhailov2006}. In nematic and polar active matter, asymmetric defects can spontaneously move and generate flows that sustain large-scale vortices~\cite{giomi2013, keber2014, giomi2015, shankar2018, angheluta2021, Farzan2022, lamontagna2025, radhakrishnan2025, Angheluta2026}. Similarly, motile dislocations are present in active smectics~\cite{lin2025}  and odd crystals~\cite{bililign2022}, while motile defects have been reported in driven XY~\cite{rouzaire2021} and nonreciprocal systems~\cite{rouzaire2025_1}. In pulsating active matter, it remains to elucidate how defects move despite the absence of any macroscopic flows and microscopic self-propulsion.

In this Letter, we reveal that defect motility emerges from a ratchet effect~\cite{Curie, Reimann2002, Reichhardt2017, Leigh2024}. The mechanochemical coupling between size oscillations and repulsion leads to asymmetric defect cores that drift under fluctuations. Our hydrodynamics captures the relevant broken spatial and time-reversal symmetries to rationalize how defect asymmetry controls motility. Overall, our symmetry arguments elucidate how repulsion and pulsation regulate the crossover between spiral and fiber-like waves, with direct implications for cardiac arrhythmia~\cite{karma2013, karma2022, Rappel2022}.


\textit{Topology of pulsating active matter}.---We consider $N$ pulsating particles in a periodic box of size $L$ with positions ${\bf r}_i$ and phases $\phi_i$. The phases effectively represent chemical cycles internal to each particle~\cite{togashi2019, boocock2023, goth2025}. Particle sizes $\sigma_i  = \sigma_0 (1+\lambda \sin \phi_i)/(1+\lambda)$ are determined by their phases $\phi_i$, where $\sigma_0=1$ is a reference size, and $\lambda<1$ sets the amplitude of deformation. Positions and phases follow overdamped Langevin dynamics~\cite{zhang2023}:
\begin{equation}\label{eq:pam_part}
\begin{aligned}
 \dot{\mathbf{r}}_i &=-\mu_r\nabla_i U\,+\,\sqrt{2 D_r} \boldsymbol{\xi}_i \ ,
 \\
 \dot{\phi}_i & =\omega - \mu_\phi \partial_{\phi_i} U + \sum_j \varepsilon_{ij}  \sin(\phi_j - \phi_i) + \sqrt{2 D_\phi} \eta_i \ .
\end{aligned}
\end{equation}
The synchronization strength is $\varepsilon_{ij} = \varepsilon$ for $a_{ij}<1$, where $a_{ij} = |{\bf r}_i - {\bf r}_i|/(\sigma_i + \sigma_j)$, and $\varepsilon_{ij}=0$ otherwise. The drive $\omega$ favors periodic pulsation of sizes, while $(\mu_r,\mu_\phi)$ and $(D_r,D_\phi)$ are the mobility and diffusion coefficients, respectively. The Gaussian white noise terms $({\bm\xi}_i,\eta_i)$ are uncorrelated with zero mean.

The repulsive potential $U = \frac 1 2 \sum_{i,j} U_0 (a_{ij}^{-12} - 2 a_{ij}^{-6} + 1 )$ has a cut-off at $a_{ij}=1$. The dependence of $U$ on positions ${\bf r}_i$ and phases $\phi_i$ yields displacement ($-\nabla_i U$) and deformation ($-\partial_{\phi_i} U$) forces when particles overlap. This mechanochemical coupling is typically absent when modelling contractile tissues as excitable media~\cite{barkley1995, cameron2012, pravdin2023, roth2001, grill1995}, and implies that $U$ is not invariant under an arbitrary rotation of phases $\phi_i$. In what follows, we reveal that the broken invariance induced by mechanochemical coupling plays a crucial role in regulating defect motility.

The phase diagram features three homogeneous states: disorder (no synchronization), cycles (synchronization with phase current), and arrest (synchronization without phase current)~\cite{zhang2023}. In between the regimes of cycles and arrest, a dynamical instability leads to deformation waves that propagate coordinated changes in phases $\phi_i$ with negligible displacement of positions ${\bf r}_i$. Topological defects of opposite charges $ \frac{1}{2\pi} \oint d\phi = \pm 1$ rotate clockwise/counterclowkise (depending on their charges), and are connected by pairs through waves [Figs.~\ref{fig1}(a-b)]; the connectivity of these pairs changes whenever waves collide. In what follows, we examine in detail how waves self-organize into various patterns associated with distinct defect statistics.

Increasing the density $\rho=N/L^2$, we observe that {\em spirals}, where defects are connected through long waves, turn into {\em fibers} with a shorter distance between connected defects [Figs.~\ref{fig2}(a-d)]. Similar spirals can be found in other models, such as reaction-diffusion dynamics described by the complex Ginzburg-landau equation (CGLE~\cite{Aranson2002}). At small $\rho$, neglecting the repulsive term $\partial_{\phi_i} U$ in the phase dynamics [Eq.~\eqref{eq:pam_part}] yields a noisy Kuramoto model~\cite{Ritort2005} (analogously, a driven XY model~\cite{rouzaire2025_2}) that maps into CGLE. In contrast, the fibers do not admit any straightforward analogy with patterns of CGLE. Indeed, the term $\partial_{\phi_i} U$ breaks the rotational invariance of the phase dynamics, so that the phases arrest and waves cannot propagate if $\rho$ exceeds a threshold $\rho_c$~\cite{sm}. At intermediate $\rho$, the emergence of waves stems from local excitations nucleating on an arrested background.

Defect displacement drastically changes between spirals and fibers. The mean-squared average $\langle \Delta x^2 \rangle$ of  displacement $\Delta x(t) = |{\bf x}(t+t_o) - {\bf x}(t_o)|$ shows that defects move slower (i.e., smaller $\langle \Delta x^2 \rangle$ at large $t$) and rotate faster (i.e., oscillating $\langle \Delta x^2 \rangle$ at small $t$) for large $1 - \rho/\rho_c$ [Fig.~\ref{fig2}(e)]. In other words, spirals are associated with slow-moving, fast-rotating defects, whereas fibers come with fast-moving, slow-rotating defects.

The distribution $P(v)$ of defect velocity $v(t) = \Delta x(t) / t $, normalized as $\int_0^\infty v \, P(v) d v = 1$ with isotropic symmetry, exhibits peaks that signal defect motility [Fig.~\ref{fig2}(f)]; we measure $v(t)$ for a time interval $t$ smaller than the typical time for defects to complete a rotation (that corresponds to the first oscillation in $\langle \Delta x^2\rangle$). At high $\rho$, defects hop along a lattice formed by the particle close packing, so $P(v)$ features several peaks corresponding to discrete jumps~\cite{sm}. For comparison, the velocity statistics of inertial Brownian particles (without motility, with mass $m$ and temperature $T$) follows the Boltzmann weight $P(v)\sim e^{-m v^2/(2 T)}$ centered at $v=0$. Here, motile defects stabilize despite the absence of any self-propulsion in the microscopic dynamics [Eq.~\eqref{eq:pam_part}]: instead, defect motility stems from an emergent mechanism associated with pattern formation.

Defect displacement has been related to shape asymmetry in various types of active matter~\cite{shankar2022}. Here, we characterize such an asymmetry by considering the rotating phase profile $\phi(\theta,t)$ as a function of the polar angle $\theta$ [Fig.~\ref{fig1}(b)]~\cite{sm}. In the frame rotating at frequency $\Omega$, the asymmetry of the profile $\phi(\theta-\Omega t)$ increases with $\rho$ [Fig.~\ref{fig2}(g)]: defect cores are more symmetric for spirals than for fibers. In fact, the asymmetry factor
\begin{equation}\label{eq:chi}
  \chi[\phi] = \frac{1}{2\pi} \int_0^{2\pi} |\phi(\theta) - \theta|\, d\theta
\end{equation}
correlates with the peak value of velocity distribution $v_p = {\rm argmax} P(v)$ [Fig.~\ref{fig2}(h)], showing that $\chi$ is a reliable predictor of defect motility.

In short, the topology of deformation waves obeys a crossover between spirals and fibers with distinct defects. Defect charge sets wave rotation, and the asymmetry factor $\chi$ suffices to distinguish between fast-moving, slow-rotating and slow-moving, fast-rotating defects. In what follows, we develop a hydrodynamic theory that helps rationalize how to control motility through asymmetry.


\textit{Broken invariance in fluctuating hydrodynamics}.---We coarse-grain our microscopic dynamics to study pattern and defect formation through a large-wavelength description. Specifically, our aim is to obtain a closed dynamics for the complex field $A({\bf r},t)$ that quantifies the local degree of phase synchronization:
\begin{equation}\label{eq:A}
	A({\bf r},t) = \frac 1 \rho \sum_j e^{i \phi_j(t)} \delta({\bf r}-{\bf r}_j(t)) = R \, e^{i\psi} \ .
\end{equation}
For simplicity, we discard interactions in the position dynamics, so that $\rho$ remains constant. We also approximate repulsion in the phase dynamics as an external field that favors small size: $\partial_{\phi_i}U  \approx h(\rho) \cos \phi_i$, where $h(\rho)=\rho^6$. We demonstrate below that these approximations enable a systematic hydrodynamic derivation that retains the essential features at play in defect motility.

Stochastic calculus~\cite{dean, Marchetti2018} yields the dynamics of the angular moments $f_n({\bf r},t) = \int e^{in\phi} P(\phi,{\bf r},t) d\phi$ derived from the empirical distribution $P(\phi,{\bf r},t) = \sum_j \delta(\phi-\phi_j(t)) \delta({\bf r}-{\bf r}_j(t))$:
\begin{equation}\label{eq:dyn_fn}
\begin{aligned}
    \partial_t f_n &= (D_r \nabla^2 + in\omega - n^2 D_\phi)f_n - (i n/2) h(\rho) (f_{n+1}+f_{n-1})
    \\
    &\quad + (n\varepsilon/2) (f_{1}f_{n-1}-f_{-1} f_{n+1}) + \sqrt{2 D_\phi} \,\Lambda_n \ .
\end{aligned}
\end{equation}
The noise $\Lambda_n=u_n+iv_n$ is Gaussian with correlations
\begin{equation}\label{eq:correlations}
\begin{aligned}
	&\langle (u_n,v_n)({\bf r},t)(u_n,v_n)^T({\bf r}',t') \rangle = {\mathbb C}_n \, \delta({\bf r}-{\bf r}')\delta(t-t') \ ,
	\\
	&{\mathbb C}_n =\frac{1}{2}
	\begin{pmatrix}
	f_0-{\rm Re}[f_{2n}] & -{\rm Im}[f_{2n}] \\
	-{\rm Im}[f_{2n}] & f_0+{\rm Re} [f_{2n}]
	\end{pmatrix} \ .
\end{aligned}
\end{equation}
In practice, we generate $(u_n,v_n)$ with two Gaussian white noises using a Cholesky decomposition of ${\mathbb C}_n$~\cite{sm}. The hierarchy in Eq.~\eqref{eq:dyn_fn} needs to be combined with a specific closure to deduce the dynamics of $A=f_1/\rho$.

\begin{figure}
\centering
	\includegraphics[width=0.5\textwidth]{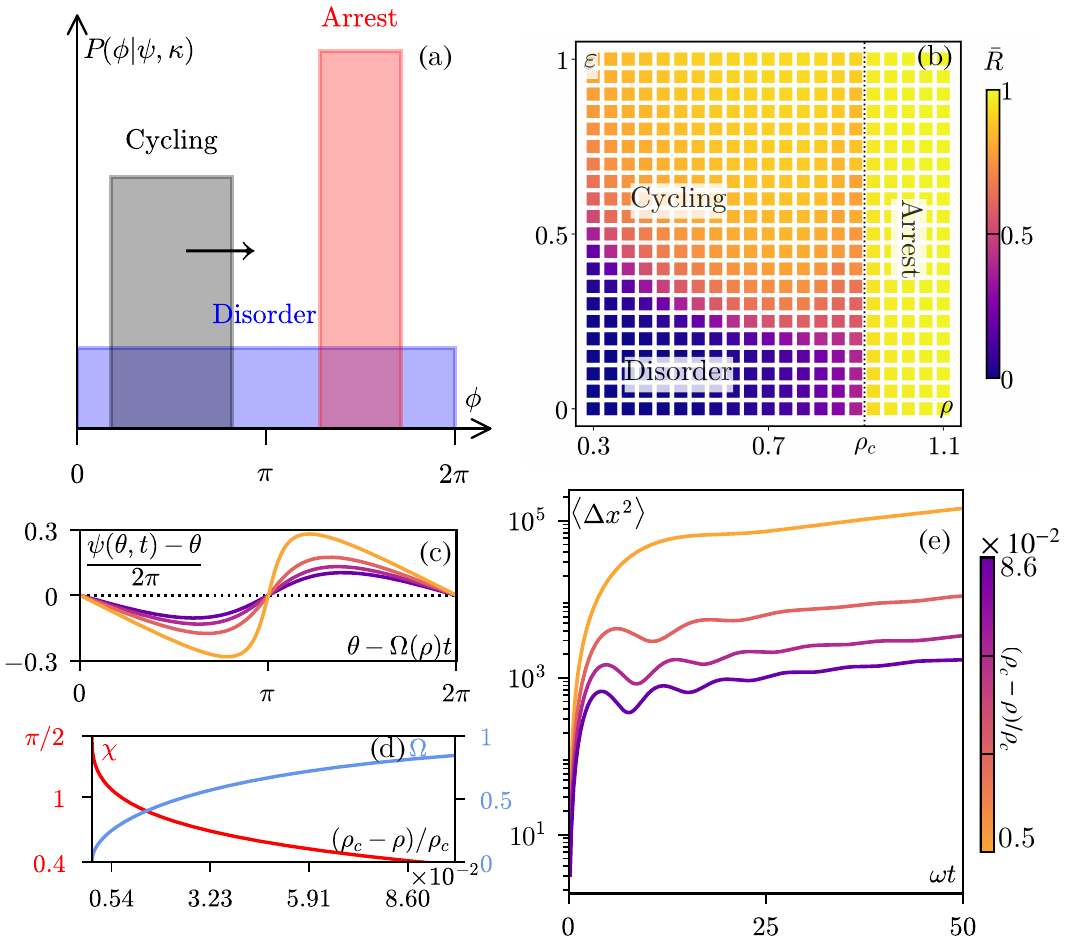}
	\caption{Fluctuating hydrodynamics [Eq.~\eqref{eq:hydro}]. (a)~Schematic of the ansatz distribution $P_{\rm ans}(\phi|\psi,\sigma)$ [Eq.~\eqref{eq:ans}].
	(b)~Phase diagram: $\bar R = \frac 1 V \int_V R({\bf r}) d{\bf r}$ [Eq.~\eqref{eq:A}] as a function of density $\rho$ and synchronization $\varepsilon$.
	(c)~Analytical defect profile $\psi(\theta,t)$ [Eq.~\eqref{eq:psi}] in the co-rotating frame $\theta-\Omega(\rho)t$; colors refer to $\rho$ values.
	(d)~Corresponding asymmetry factor $\chi[\psi]$ [Eq.~\eqref{eq:chi}] and rotational frequency $\Omega$ [Eq.~\eqref{eq:omega}].
	(e)~Mean-squared displacement $\langle\Delta x^2\rangle$ in effective defect dynamics [Eq.~\eqref{eq:dyn_def}].  Parameters in Ref.~\cite{sm}.}
\label{fig3} 
\end{figure}

Some closures in coarse-graining active models involve an adiabatic elimination of higher-order moments that is valid only close to disorder~\cite{Cates2015, Chate2020}. When deploying this strategy to pulsating active matter, it proved inadequate to properly capture the emergence of motile defects~\cite{zhang2023}. Instead, inspired by~\cite{Archer2009, Manacorda2017, Manacorda2025, Zhang2025}, we use the closure $P(\phi,{\bf r},t)\approx P_{\rm ans}(\phi \,|\, \psi({\bf r},t), \kappa({\bf r},t))$ with the ansatz
\begin{equation}\label{eq:ans}
	P_{\rm ans}(\phi \,|\, \psi, \kappa) = (\rho/\kappa) \, H(\kappa/2 - |\psi - \phi| ) \ ,
\end{equation}
where $H(x>0)=1$ and $H(x<0)=0$, leading to
\begin{equation}\label{eq:mom}
	f_n({\bf r},t) = \rho \, e^{in\psi({\bf r},t)}\sinc \left(\dfrac{n\kappa({\bf r},t)}{2}\right) \ .
\end{equation}
The ansatz $P_{\rm ans}$ parametrizes the distribution in terms of mean $\psi$ and variance $\kappa$ to distinguish the three homogeneous states: disorder ($\kappa=2\pi$), cycles ($\kappa<2\pi$ and $\dot\psi\neq 0$), and arrest ($\kappa<2\pi$ and $\dot\psi=0$) [Fig.~\ref{fig3}(a)]. In general, $(\psi,\kappa)$ are space-dependent fields whose realizations determine all the angular moments $f_n = f_n(\psi, \kappa)$ [Eq.~\eqref{eq:mom}]. Our closure enforces that $f_n$ is determined by $f_1$ (a similar strategy is used for space-independent oscillators~\cite{Antonsen2008, Strogatz2008, cestnik2022, Buendia2025, gupta2025}) under the constraint $|f_n|<\rho$ ensuring stability of noise correlations: $\det {\mathbb C}_n > 0$ [Eq.~\eqref{eq:correlations}].

The expressions $f_1 = f_1(\psi, \kappa)$ and $f_2 = f_2(\psi, \kappa)$ define a closed relation between the first and second moments: $f_2 = f_2(f_1)$. The dynamics for $A$ then directly follows from Eq.~\eqref{eq:dyn_fn} in the case $n=1$:
\begin{equation}\label{eq:hydro}
\begin{aligned}
    \partial_t A &= D_r\nabla^2A + F(A) + G(A) + \sqrt{2D_\phi} \, (\Lambda_1/\rho) \ ,
\end{aligned}
\end{equation}
where
\begin{equation}
\begin{aligned}
    F(A) &= (i\omega + \varepsilon\rho/2 - D_\phi) A - (\varepsilon/2) A^* f_2(A) \ ,
    \\
    G(A) &= (-i/2)\,h(\rho) \big[ 1 +  f_2(A)/\rho \big] \ ,
\end{aligned}
\end{equation}
and $A^*$ is the complex conjugate of $A$. Under arbitrary rotation $\Phi$ of the complex phase, $F(A)$ is symmetric as in CGLE~\cite{Aranson2002}, while $G(A)$ is not: $F(A e^{i \Phi})= e^{i \Phi} F(A)$, and $G(A e^{i \Phi})\neq e^{i \Phi} G(A)$, where we have used $f_2(A e^{i \Phi}) = e^{i2\Phi} f_2(A)$. Since $G(A)$ is proportional to the field $h(\rho)$ that embodies microscopic repulsion, this term describes how repulsion regulates the deviation of our hydrodynamics from that of rotation-invariant synchronization theories. Terms breaking rotational invariance have also been reported in the hydrodynamics of Ref.~\cite{zhang2023}, albeit with a series of caveats. We discuss below how our hydrodynamics solve these issues, and open the door to studying defect motility.

Our hydrodynamic phase diagram [Fig.~\ref{fig3}(b)] features the three homogeneous states (disorder, cycles, arrest) as in microscopics~\cite{zhang2023}. These phases bare analogies with recent theories~\cite{chatzittofi2023, martin2025, Blom2025} at mean-field level, namely without space and noise. In the ordered state [$A\approx e^{i\psi}$, Eq.~\eqref{eq:A}], arrest emerges above the critical density $\rho_c$ defined by $h(\rho_c)=\omega$~\cite{sm}. We do not observe the spurious counterclockwise cycling reported in Ref.~\cite{zhang2023}, which was recognized as an artifact of the adiabatic elimination. Between arrest and cycles, a regime of instability leads motile defects to nucleate and propagate in the arrested background [Figs.~\ref{fig1}(c-d)]. As in microscopics [Figs.~\ref{fig1}(a-b)], defect profiles are strongly asymmetric, and defect pairs are connected by fibers; Ref.~\cite{zhang2023} missed such asymmetric defects at the hydrodynamic level.

Fluctuations play an essential role in stabilizing hydrodynamic patterns. Homogeneous states are linearly stable~\cite{sm}, yet the presence of the noise term $\Lambda_1$ [Eq.~\eqref{eq:hydro}] destabilizes homogeneous profiles and selects defect profiles in our numerics. The success of our hydrodynamics stems from carefully deriving (and simulating) the multiplicative noise [Eq.~\eqref{eq:correlations}] consistently with our closure [Eq.~\eqref{eq:mom}]. In contrast, the additive noise in the hydrodynamics of Ref.~\cite{zhang2023} is inadequate to capture the nucleation of asymmetric defects in an arrested background.

In short, our fluctuating hydrodynamics captures emergent motile defects with  asymmetric profiles, thus going beyond the hydrodynamic description in Ref.~\cite{zhang2023}. We now exploit this revised hydrodynamics to examine the mechanism at play in defect motility.


\textit{Ratchet effect controls emergent defect motility}.---The emergence of motile defects is an instance of the ratchet effect~\cite{Curie, Reimann2002, Reichhardt2017, Leigh2024}: it stems from simultaneously breaking the spatial symmetry of defect profiles, as seen in other active defects~\cite{shankar2022}, and the time-reversal symmetry of the dynamics~\cite{Nardini2016, Nardini2017, Jack2022}, here due to pulsation. In what follows, we leverage our hydrodynamics to explicitly relate defect asymmetry and emergent motility.

We proceed with a series of approximations to derive the profile of the phase $\psi$ [Eq.~\eqref{eq:A}] around defects. Defects evolve in an ordered background, so we consider the solution $A\approx e^{i\psi}$ in our hydrodynamics [Eq.~\eqref{eq:hydro}]:
\begin{equation}
	\partial_t \psi = \dfrac{D_r}{r^2} \partial_{\theta\theta} \psi + \omega  - h(\rho) \cos\psi \ ,
\end{equation}
where $(r,\theta)$ are the polar coordinates in the co-moving frame [Fig.~\ref{fig1}], and we have assumed that $\psi$ depends only on $\theta$ close to defect cores. Rotating solutions exist at small density $\rho<\rho_c$, namely for high pulsation $\omega>h(\rho)$. In the frame rotating at frequency $\Omega(\rho)$, we deduce the defect profile in the regime $r \gg \sqrt{D_r/\Omega(\rho)}$~\cite{sm}:
\begin{equation}\label{eq:psi}
	\psi(\theta, t) = 2 \, {\rm arctan} \left[ \tan \left( \frac{\theta - \Omega(\rho) t}{2}\right) \frac{\Omega(\rho)}{\omega + h(\rho)} \right] \ ,
\end{equation}
where the rotational frequency reads
\begin{equation}\label{eq:omega}
  \Omega(\rho) = q \sqrt{ \omega^2 - h^2(\rho)} \ ,
\end{equation}
for a defect charge $q=\pm 1$. As $\rho$ gets closer to the arrested value $\rho_c$, namely $\omega$ approaches $h(\rho)$, we observe the same trend as in microscopics [Fig.~\ref{fig2}].

First, the profile $\psi$ gets more asymmetric [Fig.~\ref{fig3}(c)], as shown by $\chi[\psi]$ [Eq.~\eqref{eq:chi}][Fig.~\ref{fig3}(d)], which in turn increases defect motility by virtue of the racthet effect~\cite{Curie, Reimann2002, Reichhardt2017, Leigh2024}. Second, the frequency $\Omega$ decreases [Fig.~\ref{fig3}(d)]; the divergence of the period $T=2\pi/\Omega$ [Fig.~\ref{fig3}(d)] close to arrest, namely for $\omega = h(\rho)$, is common to saddle-node-infinite-period (SNIPER) bifurcations~\cite{martin2025, Blom2025}. In short, our hydrodynamics accommodates defects that rotate slower and move faster with density, as in microscopics.

We now offer an effective defect dynamics informed by the profile $\psi$ [Eq.~\eqref{eq:psi}] that captures the displacement statistics reported in microscopics [Fig.~\ref{fig2}]. The displacement results from the interplay between emergent motility (driven by asymmetric cores), intrinsic defect rotation, and defect interactions. Therefore, we map our defects into chiral motile particles~\cite{liebchen2017} with position $\bf x$ and orientation $\hat{\bf e}(\theta) = (\cos \theta, \sin \theta)$:
\begin{equation}\label{eq:dyn_def}
	\dot {\bf x} = \nu(\rho) \, \hat{\bf e}(\theta) + \sqrt{2D_t}\, \boldsymbol{\xi} \ ,
	\quad
	\dot \theta = \Omega(\rho) + \sqrt{2/\tau} \, \eta \ ,
\end{equation}
where $\nu$ is the motility, $\Omega$ the chirality, $D_t$ the translational diffusion coefficient, and $\tau$ the orientation persistence~\cite{Marchetti2018}. We map $\Omega$ into the rotational frequency of the asymmetric profile [Eq.~\eqref{eq:omega}]. $\boldsymbol{\xi}$ and $\eta$ are independent, zero-mean, unit-variance, Gaussian white noises. Defect motility emerges at one-body level due to asymmetric cores; in contrast, symmetric defects in CGLE move only through interactions~\cite{aranson1998, brito2003}. We expect defect interactions to renormalize $\nu(\rho)$.

In microscopics, the numerical evaluation of peak velocity $v_p(\rho)$ and defect asymmetry $\chi(\rho)$ [Fig.~\ref{fig2}(h)] provides the relation $v_p = v_p(\chi)$. We leverage this relation to evaluate $\nu = \nu(\chi)$ in terms of the asymmetry $\chi[\psi]$ of the hydrodynamic defect profile $\psi$ [Eq.~\eqref{eq:psi}]. This procedure yields the explicit dependence $\nu(\rho)$~\cite{sm}. The corresponding mean-squared displacement $\langle \Delta x^2 \rangle$ for the effective dynamics [Eq.~\eqref{eq:dyn_def}] captures the same crossover as in microscopics [Fig.~\ref{fig2}(e)]: defects change from slow-moving, fast-rotating to fast-moving, slow-rotating dynamics as $\rho$ increases [Fig.~\ref{fig3}(e)]. In fact, slow-rotating defects move faster with $\rho$ because of both larger motility $\nu(\rho)$ and smaller chirality $\Omega(\rho)$.

In short, the ratchet effect relates asymmetric profiles with emergent motility of defects. The corresponding defect dynamics features a crossover between two regimes of defect displacement that mirrors the change from spiral to fiber-like waves in microscopics.


\textit{Discussion}.---In pulsating active matter, the coupling between phases and positions is essential: it yields forces that break rotational invariance, which in turn stabilize patterns and defects distinct from the standard phenomenology of rotation-invariant synchronization theories~\cite{Aranson2002, Ritort2005}. Specifically, asymmetric defects regulate the self-organization of deformation waves into steady spirals and fiber-like patterns that bear analogies with the physics of cardiac tissues~\cite{karma2013, karma2022, Rappel2022}.

We have revealed that the asymmetric profile and the displacement statistics of defects hold the signature of the crossover between spirals and fibers. Using a hydrodynamic description, we have unveiled the broken symmetries that regulate this crossover through a ratchet effect. Our approach uncovers the key role of large-scale fluctuations: it destabilizes order by nucleating defects that behave as motile chiral particles. The closure of our coarse-graining can inspire novel hydrodynamic derivations in other classes of active matter~\cite{Cates2015, Chate2020, fruchart2026}.

The ability to predict and control defect dynamics has broad implications. Control theory offers a roadmap to manipulate nonequilibrium patterns~\cite{shankar2022drops, Davis2025, fbgp-qpvv, soriani2025, alvarado2025, krishnan2024} and guide strategies for steering defects towards desired configurations~\cite{Norton2020, Shankar2024, ghosh2024, ghosh2025, geerds2025}. A derivation of defect interactions, inspired by recent works~\cite{skogvoll2023, romano2023, romano2024, rana2024, lamontagna2025}, would be a fruitful starting point for a control framework in pulsating active matter.

\acknowledgments{We acknowledge insightful discussions with Francesco Serafin and Yiwei Zhang. This project has received funding from the European Union’s Horizon Europe research and innovation programme under the Marie Sk\l{}odowska-Curie grant agreement No 101056825 (NewGenActive), and from the Luxembourg National Research Fund (FNR), grant references 17962137 and 14389168. A.M. acknowledges financial support from Grant No. 2022HNW5YL MOCA funded by the Ministero dell’Università e della Ricerca PRIN2022 program.}


\bibliography{references-df}

\begin{thebibliography}{100}%
\makeatletter
\providecommand \@ifxundefined [1]{%
 \@ifx{#1\undefined}
}%
\providecommand \@ifnum [1]{%
 \ifnum #1\expandafter \@firstoftwo
 \else \expandafter \@secondoftwo
 \fi
}%
\providecommand \@ifx [1]{%
 \ifx #1\expandafter \@firstoftwo
 \else \expandafter \@secondoftwo
 \fi
}%
\providecommand \natexlab [1]{#1}%
\providecommand \enquote  [1]{``#1''}%
\providecommand \bibnamefont  [1]{#1}%
\providecommand \bibfnamefont [1]{#1}%
\providecommand \citenamefont [1]{#1}%
\providecommand \href@noop [0]{\@secondoftwo}%
\providecommand \href [0]{\begingroup \@sanitize@url \@href}%
\providecommand \@href[1]{\@@startlink{#1}\@@href}%
\providecommand \@@href[1]{\endgroup#1\@@endlink}%
\providecommand \@sanitize@url [0]{\catcode `\\12\catcode `\$12\catcode
  `\&12\catcode `\#12\catcode `\^12\catcode `\_12\catcode `\%12\relax}%
\providecommand \@@startlink[1]{}%
\providecommand \@@endlink[0]{}%
\providecommand \url  [0]{\begingroup\@sanitize@url \@url }%
\providecommand \@url [1]{\endgroup\@href {#1}{\urlprefix }}%
\providecommand \urlprefix  [0]{URL }%
\providecommand \Eprint [0]{\href }%
\providecommand \doibase [0]{https://doi.org/}%
\providecommand \selectlanguage [0]{\@gobble}%
\providecommand \bibinfo  [0]{\@secondoftwo}%
\providecommand \bibfield  [0]{\@secondoftwo}%
\providecommand \translation [1]{[#1]}%
\providecommand \BibitemOpen [0]{}%
\providecommand \bibitemStop [0]{}%
\providecommand \bibitemNoStop [0]{.\EOS\space}%
\providecommand \EOS [0]{\spacefactor3000\relax}%
\providecommand \BibitemShut  [1]{\csname bibitem#1\endcsname}%
\let\auto@bib@innerbib\@empty
\bibitem [{\citenamefont {Tjhung}\ and\ \citenamefont
  {Berthier}(2017)}]{Tjhung2017}%
  \BibitemOpen
  \bibfield  {author} {\bibinfo {author} {\bibfnamefont {E.}~\bibnamefont
  {Tjhung}}\ and\ \bibinfo {author} {\bibfnamefont {L.}~\bibnamefont
  {Berthier}},\ }\bibfield  {title} {\bibinfo {title} {Discontinuous
  fluidization transition in time-correlated assemblies of actively deforming
  particles},\ }\href {https://doi.org/10.1103/PhysRevE.96.050601} {\bibfield
  {journal} {\bibinfo  {journal} {Phys. Rev. E}\ }\textbf {\bibinfo {volume}
  {96}},\ \bibinfo {pages} {050601(R)} (\bibinfo {year} {2017})}\BibitemShut
  {NoStop}%
\bibitem [{\citenamefont {Koyano}\ \emph {et~al.}(2020)\citenamefont {Koyano},
  \citenamefont {Kitahata},\ and\ \citenamefont {Mikhailov}}]{Koyano2019}%
  \BibitemOpen
  \bibfield  {author} {\bibinfo {author} {\bibfnamefont {Y.}~\bibnamefont
  {Koyano}}, \bibinfo {author} {\bibfnamefont {H.}~\bibnamefont {Kitahata}},\
  and\ \bibinfo {author} {\bibfnamefont {A.~S.}\ \bibnamefont {Mikhailov}},\
  }\bibfield  {title} {\bibinfo {title} {Diffusion in crowded colloids of
  particles cyclically changing their shapes},\ }\href
  {https://doi.org/10.1209/0295-5075/128/40003} {\bibfield  {journal} {\bibinfo
   {journal} {Europhys. Lett.}\ }\textbf {\bibinfo {volume} {128}},\ \bibinfo
  {pages} {40003} (\bibinfo {year} {2020})}\BibitemShut {NoStop}%
\bibitem [{\citenamefont {Togashi}(2019)}]{togashi2019}%
  \BibitemOpen
  \bibfield  {author} {\bibinfo {author} {\bibfnamefont {Y.}~\bibnamefont
  {Togashi}},\ }\bibfield  {title} {\bibinfo {title} {Modeling of
  nanomachine/micromachine crowds: Interplay between the internal state and
  surroundings},\ }\href {https://doi.org/10.1021/acs.jpcb.8b10633} {\bibfield
  {journal} {\bibinfo  {journal} {J. Phys. Chem. B}\ }\textbf {\bibinfo
  {volume} {123}},\ \bibinfo {pages} {1481} (\bibinfo {year}
  {2019})}\BibitemShut {NoStop}%
\bibitem [{\citenamefont {Parisi}\ \emph {et~al.}(2023)\citenamefont {Parisi},
  \citenamefont {Wiebke}, \citenamefont {Mandl},\ and\ \citenamefont
  {Textor}}]{parisi2023}%
  \BibitemOpen
  \bibfield  {author} {\bibinfo {author} {\bibfnamefont {D.~R.}\ \bibnamefont
  {Parisi}}, \bibinfo {author} {\bibfnamefont {L.~E.}\ \bibnamefont {Wiebke}},
  \bibinfo {author} {\bibfnamefont {J.~N.}\ \bibnamefont {Mandl}},\ and\
  \bibinfo {author} {\bibfnamefont {J.}~\bibnamefont {Textor}},\ }\bibfield
  {title} {\bibinfo {title} {Flow rate resonance of actively deforming
  particles},\ }\href {https://doi.org/10.1038/s41598-023-36182-5} {\bibfield
  {journal} {\bibinfo  {journal} {Sci. Rep.}\ }\textbf {\bibinfo {volume}
  {13}},\ \bibinfo {pages} {9455} (\bibinfo {year} {2023})}\BibitemShut
  {NoStop}%
\bibitem [{\citenamefont {Li}\ \emph {et~al.}(2025)\citenamefont {Li},
  \citenamefont {Lei},\ and\ \citenamefont {qiang Ma}}]{li2024fluidization}%
  \BibitemOpen
  \bibfield  {author} {\bibinfo {author} {\bibfnamefont {Z.-Q.}\ \bibnamefont
  {Li}}, \bibinfo {author} {\bibfnamefont {Q.-L.}\ \bibnamefont {Lei}},\ and\
  \bibinfo {author} {\bibfnamefont {Y.}~\bibnamefont {qiang Ma}},\ }\bibfield
  {title} {\bibinfo {title} {Fluidization and anomalous density fluctuations in
  2d voronoi celepithelial tissues with pulsating activity},\ }\href
  {https://doi.org/10.1073/pnas.2421518122} {\bibfield  {journal} {\bibinfo
  {journal} {Proc. Natl. Acad. Sci. USA}\ }\textbf {\bibinfo {volume} {122}},\
  \bibinfo {pages} {e2421518122} (\bibinfo {year} {2025})}\BibitemShut
  {NoStop}%
\bibitem [{\citenamefont {Zhang}\ and\ \citenamefont
  {Fodor}(2023)}]{zhang2023}%
  \BibitemOpen
  \bibfield  {author} {\bibinfo {author} {\bibfnamefont {Y.}~\bibnamefont
  {Zhang}}\ and\ \bibinfo {author} {\bibfnamefont {E.}~\bibnamefont {Fodor}},\
  }\bibfield  {title} {\bibinfo {title} {Pulsating active matter},\ }\href
  {https://doi.org/10.1103/PhysRevLett.131.238302} {\bibfield  {journal}
  {\bibinfo  {journal} {Phys. Rev. Lett.}\ }\textbf {\bibinfo {volume} {131}},\
  \bibinfo {pages} {238302} (\bibinfo {year} {2023})}\BibitemShut {NoStop}%
\bibitem [{\citenamefont {Liu}\ \emph {et~al.}(2024)\citenamefont {Liu},
  \citenamefont {Zhu},\ and\ \citenamefont {Ai}}]{liu2024}%
  \BibitemOpen
  \bibfield  {author} {\bibinfo {author} {\bibfnamefont {W.~h.}\ \bibnamefont
  {Liu}}, \bibinfo {author} {\bibfnamefont {W.~j.}\ \bibnamefont {Zhu}},\ and\
  \bibinfo {author} {\bibfnamefont {B.~q.}\ \bibnamefont {Ai}},\ }\bibfield
  {title} {\bibinfo {title} {Collective motion of pulsating active particles in
  confined structures},\ }\href {https://doi.org/10.1088/1367-2630/ad23a5}
  {\bibfield  {journal} {\bibinfo  {journal} {New J. Phys.}\ }\textbf {\bibinfo
  {volume} {26}},\ \bibinfo {pages} {023017} (\bibinfo {year}
  {2024})}\BibitemShut {NoStop}%
\bibitem [{\citenamefont {Pi\~neros}\ and\ \citenamefont
  {Fodor}(2025)}]{pineros2024}%
  \BibitemOpen
  \bibfield  {author} {\bibinfo {author} {\bibfnamefont {W.~D.}\ \bibnamefont
  {Pi\~neros}}\ and\ \bibinfo {author} {\bibfnamefont {E.}~\bibnamefont
  {Fodor}},\ }\bibfield  {title} {\bibinfo {title} {Biased ensembles of
  pulsating active matter},\ }\href
  {https://doi.org/10.1103/PhysRevLett.134.038301} {\bibfield  {journal}
  {\bibinfo  {journal} {Phys. Rev. Lett.}\ }\textbf {\bibinfo {volume} {134}},\
  \bibinfo {pages} {038301} (\bibinfo {year} {2025})}\BibitemShut {NoStop}%
\bibitem [{\citenamefont {Banerjee}\ \emph
  {et~al.}(2025{\natexlab{a}})\citenamefont {Banerjee}, \citenamefont
  {Desaleux}, \citenamefont {Ranft},\ and\ \citenamefont
  {Fodor}}]{tirthankar2024}%
  \BibitemOpen
  \bibfield  {author} {\bibinfo {author} {\bibfnamefont {T.}~\bibnamefont
  {Banerjee}}, \bibinfo {author} {\bibfnamefont {T.}~\bibnamefont {Desaleux}},
  \bibinfo {author} {\bibfnamefont {J.}~\bibnamefont {Ranft}},\ and\ \bibinfo
  {author} {\bibfnamefont {{\'E}.}~\bibnamefont {Fodor}},\ }\href
  {https://arxiv.org/abs/2407.19955} {\bibinfo {title} {Hydrodynamics of
  pulsating active liquids}} (\bibinfo {year} {2025}{\natexlab{a}}),\ \Eprint
  {https://arxiv.org/abs/2407.19955} {arXiv:2407.19955 [cond-mat.soft]}
  \BibitemShut {NoStop}%
\bibitem [{\citenamefont {Banerjee}\ \emph
  {et~al.}(2025{\natexlab{b}})\citenamefont {Banerjee}, \citenamefont
  {Desaleux}, \citenamefont {Ranft},\ and\ \citenamefont
  {Fodor}}]{tirthankar2025}%
  \BibitemOpen
  \bibfield  {author} {\bibinfo {author} {\bibfnamefont {T.}~\bibnamefont
  {Banerjee}}, \bibinfo {author} {\bibfnamefont {T.}~\bibnamefont {Desaleux}},
  \bibinfo {author} {\bibfnamefont {J.}~\bibnamefont {Ranft}},\ and\ \bibinfo
  {author} {\bibfnamefont {{\'E}.}~\bibnamefont {Fodor}},\ }\href
  {https://arxiv.org/abs/2509.19024} {\bibinfo {title} {Contraction waves in
  pulsating active liquids: From pacemaker to aster dynamics}} (\bibinfo {year}
  {2025}{\natexlab{b}}),\ \Eprint {https://arxiv.org/abs/2509.19024}
  {arXiv:2509.19024 [cond-mat.stat-mech]} \BibitemShut {NoStop}%
\bibitem [{\citenamefont {Göth}\ and\ \citenamefont
  {Dzubiella}(2025)}]{goth2025}%
  \BibitemOpen
  \bibfield  {author} {\bibinfo {author} {\bibfnamefont {N.}~\bibnamefont
  {Göth}}\ and\ \bibinfo {author} {\bibfnamefont {J.}~\bibnamefont
  {Dzubiella}},\ }\bibfield  {title} {\bibinfo {title} {Collective
  chemo-mechanical oscillations and cluster waves in communicating colloids},\
  }\href {https://doi.org/10.1038/s42005-025-01983-9} {\bibfield  {journal}
  {\bibinfo  {journal} {Commun. Phys.}\ }\textbf {\bibinfo {volume} {8}},\
  \bibinfo {pages} {65} (\bibinfo {year} {2025})}\BibitemShut {NoStop}%
\bibitem [{\citenamefont {Zhu}\ \emph {et~al.}(2025)\citenamefont {Zhu},
  \citenamefont {Jiang}, \citenamefont {Li},\ and\ \citenamefont
  {Ai}}]{zhu2025}%
  \BibitemOpen
  \bibfield  {author} {\bibinfo {author} {\bibfnamefont {W.-j.}\ \bibnamefont
  {Zhu}}, \bibinfo {author} {\bibfnamefont {X.-k.}\ \bibnamefont {Jiang}},
  \bibinfo {author} {\bibfnamefont {J.-j.}\ \bibnamefont {Li}},\ and\ \bibinfo
  {author} {\bibfnamefont {B.-q.}\ \bibnamefont {Ai}},\ }\bibfield  {title}
  {\bibinfo {title} {Directed transport and collective dynamics of pulsing
  particles in topological lattices},\ }\href
  {https://doi.org/10.1103/PhysRevE.111.044123} {\bibfield  {journal} {\bibinfo
   {journal} {Phys. Rev. E}\ }\textbf {\bibinfo {volume} {111}},\ \bibinfo
  {pages} {044123} (\bibinfo {year} {2025})}\BibitemShut {NoStop}%
\bibitem [{\citenamefont {Li}\ and\ \citenamefont {Zhu}(2025)}]{Zhuli2025}%
  \BibitemOpen
  \bibfield  {author} {\bibinfo {author} {\bibfnamefont {X.-l.}\ \bibnamefont
  {Li}}\ and\ \bibinfo {author} {\bibfnamefont {W.-j.}\ \bibnamefont {Zhu}},\
  }\bibfield  {title} {\bibinfo {title} {Spontaneous demixing of a binary cell
  mixture induced by self-pulsation disparity in confluent tissues},\ }\href
  {https://doi.org/10.1103/x79c-3j1y} {\bibfield  {journal} {\bibinfo
  {journal} {Phys. Rev. E}\ }\textbf {\bibinfo {volume} {112}},\ \bibinfo
  {pages} {064407} (\bibinfo {year} {2025})}\BibitemShut {NoStop}%
\bibitem [{\citenamefont {Manning}(2023)}]{manning2023}%
  \BibitemOpen
  \bibfield  {author} {\bibinfo {author} {\bibfnamefont {M.~L.}\ \bibnamefont
  {Manning}},\ }\bibfield  {title} {\bibinfo {title} {Essay: Collections of
  deformable particles present exciting challenges for soft matter and
  biological physics},\ }\href {https://doi.org/10.1103/PhysRevLett.130.130002}
  {\bibfield  {journal} {\bibinfo  {journal} {Phys. Rev. Lett.}\ }\textbf
  {\bibinfo {volume} {130}},\ \bibinfo {pages} {130002} (\bibinfo {year}
  {2023})}\BibitemShut {NoStop}%
\bibitem [{\citenamefont {Chiou}\ \emph {et~al.}(2016)\citenamefont {Chiou},
  \citenamefont {Rocks}, \citenamefont {Chen}, \citenamefont {Cho},
  \citenamefont {Merkus}, \citenamefont {Rajaratnam}, \citenamefont {Robison},
  \citenamefont {Tewari}, \citenamefont {Vogel}, \citenamefont {Majkut},
  \citenamefont {Prosser}, \citenamefont {Discher},\ and\ \citenamefont
  {Liu}}]{chiou2016}%
  \BibitemOpen
  \bibfield  {author} {\bibinfo {author} {\bibfnamefont {K.~K.}\ \bibnamefont
  {Chiou}}, \bibinfo {author} {\bibfnamefont {J.~W.}\ \bibnamefont {Rocks}},
  \bibinfo {author} {\bibfnamefont {C.~Y.}\ \bibnamefont {Chen}}, \bibinfo
  {author} {\bibfnamefont {S.}~\bibnamefont {Cho}}, \bibinfo {author}
  {\bibfnamefont {K.~E.}\ \bibnamefont {Merkus}}, \bibinfo {author}
  {\bibfnamefont {A.}~\bibnamefont {Rajaratnam}}, \bibinfo {author}
  {\bibfnamefont {P.}~\bibnamefont {Robison}}, \bibinfo {author} {\bibfnamefont
  {M.}~\bibnamefont {Tewari}}, \bibinfo {author} {\bibfnamefont
  {K.}~\bibnamefont {Vogel}}, \bibinfo {author} {\bibfnamefont {S.~F.}\
  \bibnamefont {Majkut}}, \bibinfo {author} {\bibfnamefont {B.~L.}\
  \bibnamefont {Prosser}}, \bibinfo {author} {\bibfnamefont {D.~E.}\
  \bibnamefont {Discher}},\ and\ \bibinfo {author} {\bibfnamefont {A.~J.}\
  \bibnamefont {Liu}},\ }\bibfield  {title} {\bibinfo {title} {Mechanical
  signaling coordinates the embryonic heartbeat},\ }\href
  {https://doi.org/10.1073/pnas.1520428113} {\bibfield  {journal} {\bibinfo
  {journal} {Proc. Natl. Acad. Sci. USA}\ }\textbf {\bibinfo {volume} {113}},\
  \bibinfo {pages} {8939} (\bibinfo {year} {2016})}\BibitemShut {NoStop}%
\bibitem [{\citenamefont {Ishihara}\ \emph {et~al.}(2017)\citenamefont
  {Ishihara}, \citenamefont {Marcq},\ and\ \citenamefont
  {Sugimura}}]{ishihara2017}%
  \BibitemOpen
  \bibfield  {author} {\bibinfo {author} {\bibfnamefont {S.}~\bibnamefont
  {Ishihara}}, \bibinfo {author} {\bibfnamefont {P.}~\bibnamefont {Marcq}},\
  and\ \bibinfo {author} {\bibfnamefont {K.}~\bibnamefont {Sugimura}},\
  }\bibfield  {title} {\bibinfo {title} {From cells to tissue: A continuum
  model of epithelial mechanics},\ }\href
  {https://doi.org/10.1103/PhysRevE.96.022418} {\bibfield  {journal} {\bibinfo
  {journal} {Phys. Rev. E}\ }\textbf {\bibinfo {volume} {96}},\ \bibinfo
  {pages} {022418} (\bibinfo {year} {2017})}\BibitemShut {NoStop}%
\bibitem [{\citenamefont {Oyama}\ \emph {et~al.}(2019)\citenamefont {Oyama},
  \citenamefont {Kawasaki}, \citenamefont {Mizuno},\ and\ \citenamefont
  {Ikeda}}]{Ikeda2019}%
  \BibitemOpen
  \bibfield  {author} {\bibinfo {author} {\bibfnamefont {N.}~\bibnamefont
  {Oyama}}, \bibinfo {author} {\bibfnamefont {T.}~\bibnamefont {Kawasaki}},
  \bibinfo {author} {\bibfnamefont {H.}~\bibnamefont {Mizuno}},\ and\ \bibinfo
  {author} {\bibfnamefont {A.}~\bibnamefont {Ikeda}},\ }\bibfield  {title}
  {\bibinfo {title} {Glassy dynamics of a model of bacterial cytoplasm with
  metabolic activities},\ }\href
  {https://doi.org/10.1103/PhysRevResearch.1.032038} {\bibfield  {journal}
  {\bibinfo  {journal} {Phys. Rev. Res.}\ }\textbf {\bibinfo {volume} {1}},\
  \bibinfo {pages} {032038(R)} (\bibinfo {year} {2019})}\BibitemShut {NoStop}%
\bibitem [{\citenamefont {Staddon}\ \emph {et~al.}(2022)\citenamefont
  {Staddon}, \citenamefont {Munro},\ and\ \citenamefont
  {Banerjee}}]{Staddon-PlosOne2022}%
  \BibitemOpen
  \bibfield  {author} {\bibinfo {author} {\bibfnamefont {M.~F.}\ \bibnamefont
  {Staddon}}, \bibinfo {author} {\bibfnamefont {E.~M.}\ \bibnamefont {Munro}},\
  and\ \bibinfo {author} {\bibfnamefont {S.}~\bibnamefont {Banerjee}},\
  }\bibfield  {title} {\bibinfo {title} {Pulsatile contractions and pattern
  formation in excitable actomyosin cortex},\ }\href
  {https://doi.org/10.1371/journal.pcbi.1009981} {\bibfield  {journal}
  {\bibinfo  {journal} {PLoS Comput. Biol.}\ }\textbf {\bibinfo {volume}
  {18}},\ \bibinfo {pages} {1} (\bibinfo {year} {2022})}\BibitemShut {NoStop}%
\bibitem [{\citenamefont {Boocock}\ \emph {et~al.}(2023)\citenamefont
  {Boocock}, \citenamefont {Hirashima},\ and\ \citenamefont
  {Hannezo}}]{boocock2023}%
  \BibitemOpen
  \bibfield  {author} {\bibinfo {author} {\bibfnamefont {D.}~\bibnamefont
  {Boocock}}, \bibinfo {author} {\bibfnamefont {T.}~\bibnamefont {Hirashima}},\
  and\ \bibinfo {author} {\bibfnamefont {E.}~\bibnamefont {Hannezo}},\
  }\bibfield  {title} {\bibinfo {title} {Interplay between mechanochemical
  patterning and glassy dynamics in cellular monolayers},\ }\href
  {https://doi.org/10.1103/PRXLife.1.013001} {\bibfield  {journal} {\bibinfo
  {journal} {PRX Life}\ }\textbf {\bibinfo {volume} {1}},\ \bibinfo {pages}
  {013001} (\bibinfo {year} {2023})}\BibitemShut {NoStop}%
\bibitem [{\citenamefont {P{\'e}rez-Verdugo}\ \emph {et~al.}(2024)\citenamefont
  {P{\'e}rez-Verdugo}, \citenamefont {Banks},\ and\ \citenamefont
  {Banerjee}}]{Shiladitya2024}%
  \BibitemOpen
  \bibfield  {author} {\bibinfo {author} {\bibfnamefont {F.}~\bibnamefont
  {P{\'e}rez-Verdugo}}, \bibinfo {author} {\bibfnamefont {S.}~\bibnamefont
  {Banks}},\ and\ \bibinfo {author} {\bibfnamefont {S.}~\bibnamefont
  {Banerjee}},\ }\bibfield  {title} {\bibinfo {title} {Excitable dynamics
  driven by mechanical feedback in biological tissues},\ }\href
  {https://doi.org/10.1038/s42005-024-01661-2} {\bibfield  {journal} {\bibinfo
  {journal} {Commun. Phys.}\ }\textbf {\bibinfo {volume} {7}},\ \bibinfo
  {pages} {167} (\bibinfo {year} {2024})}\BibitemShut {NoStop}%
\bibitem [{\citenamefont {Tang}\ \emph {et~al.}(2025)\citenamefont {Tang},
  \citenamefont {Nejad}, \citenamefont {Pegoraro}, \citenamefont {Mahadevan},\
  and\ \citenamefont {Guo}}]{tang2025}%
  \BibitemOpen
  \bibfield  {author} {\bibinfo {author} {\bibfnamefont {W.}~\bibnamefont
  {Tang}}, \bibinfo {author} {\bibfnamefont {M.~R.}\ \bibnamefont {Nejad}},
  \bibinfo {author} {\bibfnamefont {A.~F.}\ \bibnamefont {Pegoraro}}, \bibinfo
  {author} {\bibfnamefont {L.}~\bibnamefont {Mahadevan}},\ and\ \bibinfo
  {author} {\bibfnamefont {M.}~\bibnamefont {Guo}},\ }\href
  {https://arxiv.org/abs/2507.16772} {\bibinfo {title} {Collective synchrony in
  confluent, pulsatile epithelia}} (\bibinfo {year} {2025}),\ \Eprint
  {https://arxiv.org/abs/2507.16772} {arXiv:2507.16772 [cond-mat.soft]}
  \BibitemShut {NoStop}%
\bibitem [{\citenamefont {Koehler}\ \emph {et~al.}(2026)\citenamefont
  {Koehler}, \citenamefont {Sandaltzopoulou}, \citenamefont {Jülicher},\ and\
  \citenamefont {Brugués}}]{koehler2026}%
  \BibitemOpen
  \bibfield  {author} {\bibinfo {author} {\bibfnamefont {L.}~\bibnamefont
  {Koehler}}, \bibinfo {author} {\bibfnamefont {E.}~\bibnamefont
  {Sandaltzopoulou}}, \bibinfo {author} {\bibfnamefont {F.}~\bibnamefont
  {Jülicher}},\ and\ \bibinfo {author} {\bibfnamefont {J.}~\bibnamefont
  {Brugués}},\ }\href {https://arxiv.org/abs/2601.05907} {\bibinfo {title}
  {Flow-wave coupling synchronizes oscillations in growing active matter}}
  (\bibinfo {year} {2026}),\ \Eprint {https://arxiv.org/abs/2601.05907}
  {arXiv:2601.05907 [physics.bio-ph]} \BibitemShut {NoStop}%
\bibitem [{\citenamefont {Karma}(2013)}]{karma2013}%
  \BibitemOpen
  \bibfield  {author} {\bibinfo {author} {\bibfnamefont {A.}~\bibnamefont
  {Karma}},\ }\bibfield  {title} {\bibinfo {title} {Physics of cardiac
  arrhythmogenesis},\ }\href
  {https://doi.org/10.1146/annurev-conmatphys-020911-125112} {\bibfield
  {journal} {\bibinfo  {journal} {Annu. Rev. Condens. Matter Phys.}\ }\textbf
  {\bibinfo {volume} {4}},\ \bibinfo {pages} {313} (\bibinfo {year}
  {2013})}\BibitemShut {NoStop}%
\bibitem [{\citenamefont {Molavi~Tabrizi}\ \emph {et~al.}(2022)\citenamefont
  {Molavi~Tabrizi}, \citenamefont {Mesgarnejad}, \citenamefont {Bazzi},
  \citenamefont {Luther}, \citenamefont {Christoph},\ and\ \citenamefont
  {Karma}}]{karma2022}%
  \BibitemOpen
  \bibfield  {author} {\bibinfo {author} {\bibfnamefont {A.}~\bibnamefont
  {Molavi~Tabrizi}}, \bibinfo {author} {\bibfnamefont {A.}~\bibnamefont
  {Mesgarnejad}}, \bibinfo {author} {\bibfnamefont {M.}~\bibnamefont {Bazzi}},
  \bibinfo {author} {\bibfnamefont {S.}~\bibnamefont {Luther}}, \bibinfo
  {author} {\bibfnamefont {J.}~\bibnamefont {Christoph}},\ and\ \bibinfo
  {author} {\bibfnamefont {A.}~\bibnamefont {Karma}},\ }\bibfield  {title}
  {\bibinfo {title} {Spatiotemporal organization of electromechanical phase
  singularities during high-frequency cardiac arrhythmias},\ }\href
  {https://doi.org/10.1103/PhysRevX.12.021052} {\bibfield  {journal} {\bibinfo
  {journal} {Phys. Rev. X}\ }\textbf {\bibinfo {volume} {12}},\ \bibinfo
  {pages} {021052} (\bibinfo {year} {2022})}\BibitemShut {NoStop}%
\bibitem [{\citenamefont {Rappel}(2022)}]{Rappel2022}%
  \BibitemOpen
  \bibfield  {author} {\bibinfo {author} {\bibfnamefont {W.-J.}\ \bibnamefont
  {Rappel}},\ }\bibfield  {title} {\bibinfo {title} {The physics of heart
  rhythm disorders},\ }\href
  {https://doi.org/https://doi.org/10.1016/j.physrep.2022.06.003} {\bibfield
  {journal} {\bibinfo  {journal} {Phys. Rep.}\ }\textbf {\bibinfo {volume}
  {978}},\ \bibinfo {pages} {1} (\bibinfo {year} {2022})},\ \bibinfo {note}
  {the physics of heart rhythm disorders}\BibitemShut {NoStop}%
\bibitem [{\citenamefont {Kim}\ \emph {et~al.}(2007)\citenamefont {Kim},
  \citenamefont {Woo}, \citenamefont {min Hwang}, \citenamefont {Hong},\ and\
  \citenamefont {Lee}}]{kim2007}%
  \BibitemOpen
  \bibfield  {author} {\bibinfo {author} {\bibfnamefont {T.~Y.}\ \bibnamefont
  {Kim}}, \bibinfo {author} {\bibfnamefont {S.-J.}\ \bibnamefont {Woo}},
  \bibinfo {author} {\bibfnamefont {S.}~\bibnamefont {min Hwang}}, \bibinfo
  {author} {\bibfnamefont {J.~H.}\ \bibnamefont {Hong}},\ and\ \bibinfo
  {author} {\bibfnamefont {K.~J.}\ \bibnamefont {Lee}},\ }\bibfield  {title}
  {\bibinfo {title} {Cardiac beat-to-beat alternations driven by unusual spiral
  waves},\ }\href {https://doi.org/10.1073/pnas.0704204104} {\bibfield
  {journal} {\bibinfo  {journal} {Proc. Natl. Acad. Sci. USA}\ }\textbf
  {\bibinfo {volume} {104}},\ \bibinfo {pages} {11639} (\bibinfo {year}
  {2007})}\BibitemShut {NoStop}%
\bibitem [{\citenamefont {Bray}(1994)}]{bray1994}%
  \BibitemOpen
  \bibfield  {author} {\bibinfo {author} {\bibfnamefont {A.}~\bibnamefont
  {Bray}},\ }\bibfield  {title} {\bibinfo {title} {Theory of phase-ordering
  kinetics},\ }\href {https://doi.org/10.1080/00018739400101505} {\bibfield
  {journal} {\bibinfo  {journal} {Adv. Phys.}\ }\textbf {\bibinfo {volume}
  {43}},\ \bibinfo {pages} {357} (\bibinfo {year} {1994})}\BibitemShut
  {NoStop}%
\bibitem [{\citenamefont {Shankar}\ \emph
  {et~al.}(2022{\natexlab{a}})\citenamefont {Shankar}, \citenamefont {Souslov},
  \citenamefont {Bowick}, \citenamefont {Marchetti},\ and\ \citenamefont
  {Vitelli}}]{shankar2022}%
  \BibitemOpen
  \bibfield  {author} {\bibinfo {author} {\bibfnamefont {S.}~\bibnamefont
  {Shankar}}, \bibinfo {author} {\bibfnamefont {A.}~\bibnamefont {Souslov}},
  \bibinfo {author} {\bibfnamefont {M.~J.}\ \bibnamefont {Bowick}}, \bibinfo
  {author} {\bibfnamefont {M.~C.}\ \bibnamefont {Marchetti}},\ and\ \bibinfo
  {author} {\bibfnamefont {V.}~\bibnamefont {Vitelli}},\ }\bibfield  {title}
  {\bibinfo {title} {Topological active matter},\ }\href
  {https://doi.org/10.1038/s42254-022-00445-3} {\bibfield  {journal} {\bibinfo
  {journal} {Nat. Rev. Phys.}\ }\textbf {\bibinfo {volume} {4}},\ \bibinfo
  {pages} {380} (\bibinfo {year} {2022}{\natexlab{a}})}\BibitemShut {NoStop}%
\bibitem [{\citenamefont {Tubiana}\ \emph {et~al.}(2024)\citenamefont
  {Tubiana}, \citenamefont {Alexander}, \citenamefont {Barbensi}, \citenamefont
  {Buck}, \citenamefont {Cartwright}, \citenamefont {Chwastyk}, \citenamefont
  {Cieplak}, \citenamefont {Coluzza}, \citenamefont {Čopar}, \citenamefont
  {Craik}, \citenamefont {{Di Stefano}}, \citenamefont {Everaers},
  \citenamefont {Faísca}, \citenamefont {Ferrari}, \citenamefont {Giacometti},
  \citenamefont {Goundaroulis}, \citenamefont {Haglund}, \citenamefont {Hou},
  \citenamefont {Ilieva}, \citenamefont {Jackson}, \citenamefont {Japaridze},
  \citenamefont {Kaplan}, \citenamefont {Klotz}, \citenamefont {Li},
  \citenamefont {Likos}, \citenamefont {Locatelli}, \citenamefont
  {López-León}, \citenamefont {Machon}, \citenamefont {Micheletti},
  \citenamefont {Michieletto}, \citenamefont {Niemi}, \citenamefont {Niemyska},
  \citenamefont {Niewieczerzal}, \citenamefont {Nitti}, \citenamefont
  {Orlandini}, \citenamefont {Pasquali}, \citenamefont {Perlinska},
  \citenamefont {Podgornik}, \citenamefont {Potestio}, \citenamefont {Pugno},
  \citenamefont {Ravnik}, \citenamefont {Ricca}, \citenamefont {Rohwer},
  \citenamefont {Rosa}, \citenamefont {Smrek}, \citenamefont {Souslov},
  \citenamefont {Stasiak}, \citenamefont {Steer}, \citenamefont {Sułkowska},
  \citenamefont {Sułkowski}, \citenamefont {Sumners}, \citenamefont
  {Svaneborg}, \citenamefont {Szymczak}, \citenamefont {Tarenzi}, \citenamefont
  {Travasso}, \citenamefont {Virnau}, \citenamefont {Vlassopoulos},
  \citenamefont {Ziherl},\ and\ \citenamefont {Žumer}}]{tubiana2024}%
  \BibitemOpen
  \bibfield  {author} {\bibinfo {author} {\bibfnamefont {L.}~\bibnamefont
  {Tubiana}}, \bibinfo {author} {\bibfnamefont {G.~P.}\ \bibnamefont
  {Alexander}}, \bibinfo {author} {\bibfnamefont {A.}~\bibnamefont {Barbensi}},
  \bibinfo {author} {\bibfnamefont {D.}~\bibnamefont {Buck}}, \bibinfo {author}
  {\bibfnamefont {J.~H.}\ \bibnamefont {Cartwright}}, \bibinfo {author}
  {\bibfnamefont {M.}~\bibnamefont {Chwastyk}}, \bibinfo {author}
  {\bibfnamefont {M.}~\bibnamefont {Cieplak}}, \bibinfo {author} {\bibfnamefont
  {I.}~\bibnamefont {Coluzza}}, \bibinfo {author} {\bibfnamefont
  {S.}~\bibnamefont {Čopar}}, \bibinfo {author} {\bibfnamefont {D.~J.}\
  \bibnamefont {Craik}}, \bibinfo {author} {\bibfnamefont {M.}~\bibnamefont
  {{Di Stefano}}}, \bibinfo {author} {\bibfnamefont {R.}~\bibnamefont
  {Everaers}}, \bibinfo {author} {\bibfnamefont {P.~F.}\ \bibnamefont
  {Faísca}}, \bibinfo {author} {\bibfnamefont {F.}~\bibnamefont {Ferrari}},
  \bibinfo {author} {\bibfnamefont {A.}~\bibnamefont {Giacometti}}, \bibinfo
  {author} {\bibfnamefont {D.}~\bibnamefont {Goundaroulis}}, \bibinfo {author}
  {\bibfnamefont {E.}~\bibnamefont {Haglund}}, \bibinfo {author} {\bibfnamefont
  {Y.-M.}\ \bibnamefont {Hou}}, \bibinfo {author} {\bibfnamefont
  {N.}~\bibnamefont {Ilieva}}, \bibinfo {author} {\bibfnamefont {S.~E.}\
  \bibnamefont {Jackson}}, \bibinfo {author} {\bibfnamefont {A.}~\bibnamefont
  {Japaridze}}, \bibinfo {author} {\bibfnamefont {N.}~\bibnamefont {Kaplan}},
  \bibinfo {author} {\bibfnamefont {A.~R.}\ \bibnamefont {Klotz}}, \bibinfo
  {author} {\bibfnamefont {H.}~\bibnamefont {Li}}, \bibinfo {author}
  {\bibfnamefont {C.~N.}\ \bibnamefont {Likos}}, \bibinfo {author}
  {\bibfnamefont {E.}~\bibnamefont {Locatelli}}, \bibinfo {author}
  {\bibfnamefont {T.}~\bibnamefont {López-León}}, \bibinfo {author}
  {\bibfnamefont {T.}~\bibnamefont {Machon}}, \bibinfo {author} {\bibfnamefont
  {C.}~\bibnamefont {Micheletti}}, \bibinfo {author} {\bibfnamefont
  {D.}~\bibnamefont {Michieletto}}, \bibinfo {author} {\bibfnamefont
  {A.}~\bibnamefont {Niemi}}, \bibinfo {author} {\bibfnamefont
  {W.}~\bibnamefont {Niemyska}}, \bibinfo {author} {\bibfnamefont
  {S.}~\bibnamefont {Niewieczerzal}}, \bibinfo {author} {\bibfnamefont
  {F.}~\bibnamefont {Nitti}}, \bibinfo {author} {\bibfnamefont
  {E.}~\bibnamefont {Orlandini}}, \bibinfo {author} {\bibfnamefont
  {S.}~\bibnamefont {Pasquali}}, \bibinfo {author} {\bibfnamefont {A.~P.}\
  \bibnamefont {Perlinska}}, \bibinfo {author} {\bibfnamefont {R.}~\bibnamefont
  {Podgornik}}, \bibinfo {author} {\bibfnamefont {R.}~\bibnamefont {Potestio}},
  \bibinfo {author} {\bibfnamefont {N.~M.}\ \bibnamefont {Pugno}}, \bibinfo
  {author} {\bibfnamefont {M.}~\bibnamefont {Ravnik}}, \bibinfo {author}
  {\bibfnamefont {R.}~\bibnamefont {Ricca}}, \bibinfo {author} {\bibfnamefont
  {C.~M.}\ \bibnamefont {Rohwer}}, \bibinfo {author} {\bibfnamefont
  {A.}~\bibnamefont {Rosa}}, \bibinfo {author} {\bibfnamefont {J.}~\bibnamefont
  {Smrek}}, \bibinfo {author} {\bibfnamefont {A.}~\bibnamefont {Souslov}},
  \bibinfo {author} {\bibfnamefont {A.}~\bibnamefont {Stasiak}}, \bibinfo
  {author} {\bibfnamefont {D.}~\bibnamefont {Steer}}, \bibinfo {author}
  {\bibfnamefont {J.}~\bibnamefont {Sułkowska}}, \bibinfo {author}
  {\bibfnamefont {P.}~\bibnamefont {Sułkowski}}, \bibinfo {author}
  {\bibfnamefont {D.~W.~L.}\ \bibnamefont {Sumners}}, \bibinfo {author}
  {\bibfnamefont {C.}~\bibnamefont {Svaneborg}}, \bibinfo {author}
  {\bibfnamefont {P.}~\bibnamefont {Szymczak}}, \bibinfo {author}
  {\bibfnamefont {T.}~\bibnamefont {Tarenzi}}, \bibinfo {author} {\bibfnamefont
  {R.}~\bibnamefont {Travasso}}, \bibinfo {author} {\bibfnamefont
  {P.}~\bibnamefont {Virnau}}, \bibinfo {author} {\bibfnamefont
  {D.}~\bibnamefont {Vlassopoulos}}, \bibinfo {author} {\bibfnamefont
  {P.}~\bibnamefont {Ziherl}},\ and\ \bibinfo {author} {\bibfnamefont
  {S.}~\bibnamefont {Žumer}},\ }\bibfield  {title} {\bibinfo {title} {Topology
  in soft and biological matter},\ }\href
  {https://doi.org/https://doi.org/10.1016/j.physrep.2024.04.002} {\bibfield
  {journal} {\bibinfo  {journal} {Phys. Rep.}\ }\textbf {\bibinfo {volume}
  {1075}},\ \bibinfo {pages} {1} (\bibinfo {year} {2024})}\BibitemShut
  {NoStop}%
\bibitem [{\citenamefont {Agudo-Canalejo}\ and\ \citenamefont
  {Tang}(2025)}]{Tang2025_b}%
  \BibitemOpen
  \bibfield  {author} {\bibinfo {author} {\bibfnamefont {J.}~\bibnamefont
  {Agudo-Canalejo}}\ and\ \bibinfo {author} {\bibfnamefont {E.}~\bibnamefont
  {Tang}},\ }\bibfield  {title} {\bibinfo {title} {Topological phases in
  discrete stochastic systems},\ }\href
  {https://doi.org/10.1088/1361-6633/ae07fd} {\bibfield  {journal} {\bibinfo
  {journal} {Rep. Prog. Phys.}\ }\textbf {\bibinfo {volume} {88}},\ \bibinfo
  {pages} {102601} (\bibinfo {year} {2025})}\BibitemShut {NoStop}%
\bibitem [{\citenamefont {Barkley}(1995)}]{barkley1995}%
  \BibitemOpen
  \bibfield  {author} {\bibinfo {author} {\bibfnamefont {D.}~\bibnamefont
  {Barkley}},\ }\bibinfo {title} {Spiral meandering},\ in\ \href
  {https://doi.org/10.1007/978-94-011-1156-0_5} {\emph {\bibinfo {booktitle}
  {Chemical Waves and Patterns}}}\ (\bibinfo  {publisher} {Springer
  Netherlands},\ \bibinfo {year} {1995})\ pp.\ \bibinfo {pages}
  {163--189}\BibitemShut {NoStop}%
\bibitem [{\citenamefont {García-Ojalvo}\ and\ \citenamefont
  {Schimansky-Geier}(1999)}]{garcia1999}%
  \BibitemOpen
  \bibfield  {author} {\bibinfo {author} {\bibfnamefont {J.}~\bibnamefont
  {García-Ojalvo}}\ and\ \bibinfo {author} {\bibfnamefont {L.}~\bibnamefont
  {Schimansky-Geier}},\ }\bibfield  {title} {\bibinfo {title} {Noise-induced
  spiral dynamics in excitable media},\ }\href
  {https://doi.org/10.1209/epl/i1999-00388-9} {\bibfield  {journal} {\bibinfo
  {journal} {Europhys. Lett.}\ }\textbf {\bibinfo {volume} {47}},\ \bibinfo
  {pages} {298} (\bibinfo {year} {1999})}\BibitemShut {NoStop}%
\bibitem [{\citenamefont {Biktasheva}\ and\ \citenamefont
  {Biktashev}(2003)}]{biktashev2003}%
  \BibitemOpen
  \bibfield  {author} {\bibinfo {author} {\bibfnamefont {I.~V.}\ \bibnamefont
  {Biktasheva}}\ and\ \bibinfo {author} {\bibfnamefont {V.~N.}\ \bibnamefont
  {Biktashev}},\ }\bibfield  {title} {\bibinfo {title} {Wave-particle dualism
  of spiral waves dynamics},\ }\href
  {https://doi.org/10.1103/PhysRevE.67.026221} {\bibfield  {journal} {\bibinfo
  {journal} {Phys. Rev. E}\ }\textbf {\bibinfo {volume} {67}},\ \bibinfo
  {pages} {026221} (\bibinfo {year} {2003})}\BibitemShut {NoStop}%
\bibitem [{\citenamefont {Cameron}\ and\ \citenamefont
  {Davidsen}(2012)}]{cameron2012}%
  \BibitemOpen
  \bibfield  {author} {\bibinfo {author} {\bibfnamefont {T.}~\bibnamefont
  {Cameron}}\ and\ \bibinfo {author} {\bibfnamefont {J.}~\bibnamefont
  {Davidsen}},\ }\bibfield  {title} {\bibinfo {title} {Induced spiral motion in
  cardiac tissue due to alternans},\ }\href
  {https://doi.org/10.1103/PhysRevE.86.061908} {\bibfield  {journal} {\bibinfo
  {journal} {Phys. Rev. E}\ }\textbf {\bibinfo {volume} {86}},\ \bibinfo
  {pages} {061908} (\bibinfo {year} {2012})}\BibitemShut {NoStop}%
\bibitem [{\citenamefont {Pravdin}\ \emph {et~al.}(2023)\citenamefont
  {Pravdin}, \citenamefont {Patrakeev},\ and\ \citenamefont
  {Panfilov}}]{pravdin2023}%
  \BibitemOpen
  \bibfield  {author} {\bibinfo {author} {\bibfnamefont {S.~F.}\ \bibnamefont
  {Pravdin}}, \bibinfo {author} {\bibfnamefont {M.~A.}\ \bibnamefont
  {Patrakeev}},\ and\ \bibinfo {author} {\bibfnamefont {A.~V.}\ \bibnamefont
  {Panfilov}},\ }\bibfield  {title} {\bibinfo {title} {Meander pattern of
  spiral wave and the spatial distribution of its cycle length},\ }\href
  {https://doi.org/10.1103/PhysRevE.107.014215} {\bibfield  {journal} {\bibinfo
   {journal} {Phys. Rev. E}\ }\textbf {\bibinfo {volume} {107}},\ \bibinfo
  {pages} {014215} (\bibinfo {year} {2023})}\BibitemShut {NoStop}%
\bibitem [{\citenamefont {Roth}(2001)}]{roth2001}%
  \BibitemOpen
  \bibfield  {author} {\bibinfo {author} {\bibfnamefont {B.~J.}\ \bibnamefont
  {Roth}},\ }\bibfield  {title} {\bibinfo {title} {Meandering of spiral waves
  in anisotropic cardiac tissue},\ }\href
  {https://doi.org/https://doi.org/10.1016/S0167-2789(01)00145-2} {\bibfield
  {journal} {\bibinfo  {journal} {Physica D: Nonlinear Phenomena}\ }\textbf
  {\bibinfo {volume} {150}},\ \bibinfo {pages} {127} (\bibinfo {year}
  {2001})}\BibitemShut {NoStop}%
\bibitem [{\citenamefont {Grill}\ \emph {et~al.}(1995)\citenamefont {Grill},
  \citenamefont {Zykov},\ and\ \citenamefont {M\"uller}}]{grill1995}%
  \BibitemOpen
  \bibfield  {author} {\bibinfo {author} {\bibfnamefont {S.}~\bibnamefont
  {Grill}}, \bibinfo {author} {\bibfnamefont {V.~S.}\ \bibnamefont {Zykov}},\
  and\ \bibinfo {author} {\bibfnamefont {S.~C.}\ \bibnamefont {M\"uller}},\
  }\bibfield  {title} {\bibinfo {title} {Feedback-controlled dynamics of
  meandering spiral waves},\ }\href
  {https://doi.org/10.1103/PhysRevLett.75.3368} {\bibfield  {journal} {\bibinfo
   {journal} {Phys. Rev. Lett.}\ }\textbf {\bibinfo {volume} {75}},\ \bibinfo
  {pages} {3368} (\bibinfo {year} {1995})}\BibitemShut {NoStop}%
\bibitem [{\citenamefont {Mikhailov}\ and\ \citenamefont
  {Showalter}(2006)}]{mikhailov2006}%
  \BibitemOpen
  \bibfield  {author} {\bibinfo {author} {\bibfnamefont {A.~S.}\ \bibnamefont
  {Mikhailov}}\ and\ \bibinfo {author} {\bibfnamefont {K.}~\bibnamefont
  {Showalter}},\ }\bibfield  {title} {\bibinfo {title} {Control of waves,
  patterns and turbulence in chemical systems},\ }\href
  {https://doi.org/https://doi.org/10.1016/j.physrep.2005.11.003} {\bibfield
  {journal} {\bibinfo  {journal} {Phys. Rep.}\ }\textbf {\bibinfo {volume}
  {425}},\ \bibinfo {pages} {79} (\bibinfo {year} {2006})}\BibitemShut
  {NoStop}%
\bibitem [{\citenamefont {Giomi}\ \emph {et~al.}(2013)\citenamefont {Giomi},
  \citenamefont {Bowick}, \citenamefont {Ma},\ and\ \citenamefont
  {Marchetti}}]{giomi2013}%
  \BibitemOpen
  \bibfield  {author} {\bibinfo {author} {\bibfnamefont {L.}~\bibnamefont
  {Giomi}}, \bibinfo {author} {\bibfnamefont {M.~J.}\ \bibnamefont {Bowick}},
  \bibinfo {author} {\bibfnamefont {X.}~\bibnamefont {Ma}},\ and\ \bibinfo
  {author} {\bibfnamefont {M.~C.}\ \bibnamefont {Marchetti}},\ }\bibfield
  {title} {\bibinfo {title} {Defect annihilation and proliferation in active
  nematics},\ }\href {https://doi.org/10.1103/PhysRevLett.110.228101}
  {\bibfield  {journal} {\bibinfo  {journal} {Phys. Rev. Lett.}\ }\textbf
  {\bibinfo {volume} {110}},\ \bibinfo {pages} {228101} (\bibinfo {year}
  {2013})}\BibitemShut {NoStop}%
\bibitem [{\citenamefont {Keber}\ \emph {et~al.}(2014)\citenamefont {Keber},
  \citenamefont {Loiseau}, \citenamefont {Sanchez}, \citenamefont {DeCamp},
  \citenamefont {Giomi}, \citenamefont {Bowick}, \citenamefont {Marchetti},
  \citenamefont {Dogic},\ and\ \citenamefont {Bausch}}]{keber2014}%
  \BibitemOpen
  \bibfield  {author} {\bibinfo {author} {\bibfnamefont {F.~C.}\ \bibnamefont
  {Keber}}, \bibinfo {author} {\bibfnamefont {E.}~\bibnamefont {Loiseau}},
  \bibinfo {author} {\bibfnamefont {T.}~\bibnamefont {Sanchez}}, \bibinfo
  {author} {\bibfnamefont {S.~J.}\ \bibnamefont {DeCamp}}, \bibinfo {author}
  {\bibfnamefont {L.}~\bibnamefont {Giomi}}, \bibinfo {author} {\bibfnamefont
  {M.~J.}\ \bibnamefont {Bowick}}, \bibinfo {author} {\bibfnamefont {M.~C.}\
  \bibnamefont {Marchetti}}, \bibinfo {author} {\bibfnamefont {Z.}~\bibnamefont
  {Dogic}},\ and\ \bibinfo {author} {\bibfnamefont {A.~R.}\ \bibnamefont
  {Bausch}},\ }\bibfield  {title} {\bibinfo {title} {Topology and dynamics of
  active nematic vesicles},\ }\href {https://doi.org/10.1126/science.1254784}
  {\bibfield  {journal} {\bibinfo  {journal} {Science}\ }\textbf {\bibinfo
  {volume} {345}},\ \bibinfo {pages} {1135} (\bibinfo {year}
  {2014})}\BibitemShut {NoStop}%
\bibitem [{\citenamefont {Giomi}(2015)}]{giomi2015}%
  \BibitemOpen
  \bibfield  {author} {\bibinfo {author} {\bibfnamefont {L.}~\bibnamefont
  {Giomi}},\ }\bibfield  {title} {\bibinfo {title} {Geometry and topology of
  turbulence in active nematics},\ }\href
  {https://doi.org/10.1103/PhysRevX.5.031003} {\bibfield  {journal} {\bibinfo
  {journal} {Phys. Rev. X}\ }\textbf {\bibinfo {volume} {5}},\ \bibinfo {pages}
  {031003} (\bibinfo {year} {2015})}\BibitemShut {NoStop}%
\bibitem [{\citenamefont {Shankar}\ \emph {et~al.}(2018)\citenamefont
  {Shankar}, \citenamefont {Ramaswamy}, \citenamefont {Marchetti},\ and\
  \citenamefont {Bowick}}]{shankar2018}%
  \BibitemOpen
  \bibfield  {author} {\bibinfo {author} {\bibfnamefont {S.}~\bibnamefont
  {Shankar}}, \bibinfo {author} {\bibfnamefont {S.}~\bibnamefont {Ramaswamy}},
  \bibinfo {author} {\bibfnamefont {M.~C.}\ \bibnamefont {Marchetti}},\ and\
  \bibinfo {author} {\bibfnamefont {M.~J.}\ \bibnamefont {Bowick}},\ }\bibfield
   {title} {\bibinfo {title} {Defect unbinding in active nematics},\ }\href
  {https://doi.org/10.1103/PhysRevLett.121.108002} {\bibfield  {journal}
  {\bibinfo  {journal} {Phys. Rev. Lett.}\ }\textbf {\bibinfo {volume} {121}},\
  \bibinfo {pages} {108002} (\bibinfo {year} {2018})}\BibitemShut {NoStop}%
\bibitem [{\citenamefont {Angheluta}\ \emph {et~al.}(2021)\citenamefont
  {Angheluta}, \citenamefont {Chen}, \citenamefont {Marchetti},\ and\
  \citenamefont {Bowick}}]{angheluta2021}%
  \BibitemOpen
  \bibfield  {author} {\bibinfo {author} {\bibfnamefont {L.}~\bibnamefont
  {Angheluta}}, \bibinfo {author} {\bibfnamefont {Z.}~\bibnamefont {Chen}},
  \bibinfo {author} {\bibfnamefont {M.~C.}\ \bibnamefont {Marchetti}},\ and\
  \bibinfo {author} {\bibfnamefont {M.~J.}\ \bibnamefont {Bowick}},\ }\bibfield
   {title} {\bibinfo {title} {The role of fluid flow in the dynamics of active
  nematic defects},\ }\href {https://doi.org/10.1088/1367-2630/abe8a8}
  {\bibfield  {journal} {\bibinfo  {journal} {New J. Phys.}\ }\textbf {\bibinfo
  {volume} {23}},\ \bibinfo {pages} {033009} (\bibinfo {year}
  {2021})}\BibitemShut {NoStop}%
\bibitem [{\citenamefont {Vafa}(2022)}]{Farzan2022}%
  \BibitemOpen
  \bibfield  {author} {\bibinfo {author} {\bibfnamefont {F.}~\bibnamefont
  {Vafa}},\ }\bibfield  {title} {\bibinfo {title} {Defect dynamics in active
  polar fluids vs. active nematics},\ }\href
  {https://doi.org/10.1039/D2SM00830K} {\bibfield  {journal} {\bibinfo
  {journal} {Soft Matter}\ }\textbf {\bibinfo {volume} {18}},\ \bibinfo {pages}
  {8087} (\bibinfo {year} {2022})}\BibitemShut {NoStop}%
\bibitem [{\citenamefont {La~Montagna}\ \emph {et~al.}(2026)\citenamefont
  {La~Montagna}, \citenamefont {Burekovi\ifmmode~\acute{c}\else \'{c}\fi{}},
  \citenamefont {Maitra},\ and\ \citenamefont {Nardini}}]{lamontagna2025}%
  \BibitemOpen
  \bibfield  {author} {\bibinfo {author} {\bibfnamefont {G.~M.}\ \bibnamefont
  {La~Montagna}}, \bibinfo {author} {\bibfnamefont {S.}~\bibnamefont
  {Burekovi\ifmmode~\acute{c}\else \'{c}\fi{}}}, \bibinfo {author}
  {\bibfnamefont {A.}~\bibnamefont {Maitra}},\ and\ \bibinfo {author}
  {\bibfnamefont {C.}~\bibnamefont {Nardini}},\ }\bibfield  {title} {\bibinfo
  {title} {Shape dictates the motion of topological defects in active
  nematics},\ }\href {https://doi.org/10.1103/b9g7-h378} {\bibfield  {journal}
  {\bibinfo  {journal} {Phys. Rev. E}\ }\textbf {\bibinfo {volume} {113}},\
  \bibinfo {pages} {L053401} (\bibinfo {year} {2026})}\BibitemShut {NoStop}%
\bibitem [{\citenamefont {Radhakrishnan}\ \emph {et~al.}(2026)\citenamefont
  {Radhakrishnan}, \citenamefont {Serafin}, \citenamefont {Schmidt},\ and\
  \citenamefont {Fodor}}]{radhakrishnan2025}%
  \BibitemOpen
  \bibfield  {author} {\bibinfo {author} {\bibfnamefont {B.~N.}\ \bibnamefont
  {Radhakrishnan}}, \bibinfo {author} {\bibfnamefont {F.}~\bibnamefont
  {Serafin}}, \bibinfo {author} {\bibfnamefont {T.~L.}\ \bibnamefont
  {Schmidt}},\ and\ \bibinfo {author} {\bibfnamefont {{\'E}.}~\bibnamefont
  {Fodor}},\ }\bibfield  {title} {\bibinfo {title} {Irreversibility in scalar
  active turbulence: the role of topological defects},\ }\href
  {https://doi.org/10.1088/1367-2630/ae44c3} {\bibfield  {journal} {\bibinfo
  {journal} {New J. Phys.}\ }\textbf {\bibinfo {volume} {28}},\ \bibinfo
  {pages} {034601} (\bibinfo {year} {2026})}\BibitemShut {NoStop}%
\bibitem [{\citenamefont {Angheluta}\ \emph {et~al.}(2026)\citenamefont
  {Angheluta}, \citenamefont {Lång}, \citenamefont {Lång},\ and\
  \citenamefont {Bøe}}]{Angheluta2026}%
  \BibitemOpen
  \bibfield  {author} {\bibinfo {author} {\bibfnamefont {L.}~\bibnamefont
  {Angheluta}}, \bibinfo {author} {\bibfnamefont {A.}~\bibnamefont {Lång}},
  \bibinfo {author} {\bibfnamefont {E.}~\bibnamefont {Lång}},\ and\ \bibinfo
  {author} {\bibfnamefont {S.~O.}\ \bibnamefont {Bøe}},\ }\bibfield  {title}
  {\bibinfo {title} {Full-integer topological defects in polar active matter:
  From collective migration to tissue patterning},\ }\href
  {https://doi.org/https://doi.org/10.1146/annurev-conmatphys-031620-105420}
  {\bibfield  {journal} {\bibinfo  {journal} {Annu. Rev. Condens. Matter
  Phys.}\ }\textbf {\bibinfo {volume} {17}},\ \bibinfo {pages} {305} (\bibinfo
  {year} {2026})}\BibitemShut {NoStop}%
\bibitem [{\citenamefont {Lin}\ \emph {et~al.}(2026)\citenamefont {Lin},
  \citenamefont {J{\"u}licher}, \citenamefont {Prost},\ and\ \citenamefont
  {Rupprecht}}]{lin2025}%
  \BibitemOpen
  \bibfield  {author} {\bibinfo {author} {\bibfnamefont {S.-Z.}\ \bibnamefont
  {Lin}}, \bibinfo {author} {\bibfnamefont {F.}~\bibnamefont {J{\"u}licher}},
  \bibinfo {author} {\bibfnamefont {J.}~\bibnamefont {Prost}},\ and\ \bibinfo
  {author} {\bibfnamefont {J.-F.}\ \bibnamefont {Rupprecht}},\ }\bibfield
  {title} {\bibinfo {title} {Spontaneous flows in active smectics with
  dislocations},\ }\href {https://doi.org/10.1140/epjs/s11734-025-01904-5}
  {\bibfield  {journal} {\bibinfo  {journal} {Eur. Phys. J. Spec. Top.}\
  }\textbf {\bibinfo {volume} {235}},\ \bibinfo {pages} {157} (\bibinfo {year}
  {2026})}\BibitemShut {NoStop}%
\bibitem [{\citenamefont {Bililign}\ \emph {et~al.}(2022)\citenamefont
  {Bililign}, \citenamefont {Balboa~Usabiaga}, \citenamefont {Ganan},
  \citenamefont {Poncet}, \citenamefont {Soni}, \citenamefont {Magkiriadou},
  \citenamefont {Shelley}, \citenamefont {Bartolo},\ and\ \citenamefont
  {Irvine}}]{bililign2022}%
  \BibitemOpen
  \bibfield  {author} {\bibinfo {author} {\bibfnamefont {E.~S.}\ \bibnamefont
  {Bililign}}, \bibinfo {author} {\bibfnamefont {F.}~\bibnamefont
  {Balboa~Usabiaga}}, \bibinfo {author} {\bibfnamefont {Y.~A.}\ \bibnamefont
  {Ganan}}, \bibinfo {author} {\bibfnamefont {A.}~\bibnamefont {Poncet}},
  \bibinfo {author} {\bibfnamefont {V.}~\bibnamefont {Soni}}, \bibinfo {author}
  {\bibfnamefont {S.}~\bibnamefont {Magkiriadou}}, \bibinfo {author}
  {\bibfnamefont {M.~J.}\ \bibnamefont {Shelley}}, \bibinfo {author}
  {\bibfnamefont {D.}~\bibnamefont {Bartolo}},\ and\ \bibinfo {author}
  {\bibfnamefont {W.~T.~M.}\ \bibnamefont {Irvine}},\ }\bibfield  {title}
  {\bibinfo {title} {Motile dislocations knead odd crystals into whorls},\
  }\href {https://doi.org/10.1038/s41567-021-01429-3} {\bibfield  {journal}
  {\bibinfo  {journal} {Nat. Phys.}\ }\textbf {\bibinfo {volume} {18}},\
  \bibinfo {pages} {212} (\bibinfo {year} {2022})}\BibitemShut {NoStop}%
\bibitem [{\citenamefont {Rouzaire}\ and\ \citenamefont
  {Levis}(2021)}]{rouzaire2021}%
  \BibitemOpen
  \bibfield  {author} {\bibinfo {author} {\bibfnamefont {Y.}~\bibnamefont
  {Rouzaire}}\ and\ \bibinfo {author} {\bibfnamefont {D.}~\bibnamefont
  {Levis}},\ }\bibfield  {title} {\bibinfo {title} {Defect superdiffusion and
  unbinding in a 2d xy model of self-driven rotors},\ }\href
  {https://doi.org/10.1103/PhysRevLett.127.088004} {\bibfield  {journal}
  {\bibinfo  {journal} {Phys. Rev. Lett.}\ }\textbf {\bibinfo {volume} {127}},\
  \bibinfo {pages} {088004} (\bibinfo {year} {2021})}\BibitemShut {NoStop}%
\bibitem [{\citenamefont {Rouzaire}\ \emph
  {et~al.}(2025{\natexlab{a}})\citenamefont {Rouzaire}, \citenamefont {Pearce},
  \citenamefont {Pagonabarraga},\ and\ \citenamefont {Levis}}]{rouzaire2025_1}%
  \BibitemOpen
  \bibfield  {author} {\bibinfo {author} {\bibfnamefont {Y.}~\bibnamefont
  {Rouzaire}}, \bibinfo {author} {\bibfnamefont {D.~J.~G.}\ \bibnamefont
  {Pearce}}, \bibinfo {author} {\bibfnamefont {I.}~\bibnamefont
  {Pagonabarraga}},\ and\ \bibinfo {author} {\bibfnamefont {D.}~\bibnamefont
  {Levis}},\ }\bibfield  {title} {\bibinfo {title} {Nonreciprocal interactions
  reshape topological defect annihilation},\ }\href
  {https://doi.org/10.1103/PhysRevLett.134.167101} {\bibfield  {journal}
  {\bibinfo  {journal} {Phys. Rev. Lett.}\ }\textbf {\bibinfo {volume} {134}},\
  \bibinfo {pages} {167101} (\bibinfo {year} {2025}{\natexlab{a}})}\BibitemShut
  {NoStop}%
\bibitem [{sm()}]{sm}%
  \BibitemOpen
  \href@noop {} {}\bibinfo {note} {See Supplemental Material at [URL will be
  inserted by publisher] for details on analytical derivations and numerical
  simulations, which includes Refs.~\cite{munkres1957, Glasserman,
  cai2019}}\BibitemShut {NoStop}%
\bibitem [{\citenamefont {Curie}(1894)}]{Curie}%
  \BibitemOpen
  \bibfield  {author} {\bibinfo {author} {\bibfnamefont {P.}~\bibnamefont
  {Curie}},\ }\bibfield  {title} {\bibinfo {title} {Sur la symétrie dans les
  phénomènes physiques, symétrie d'un champ électrique et d'un champ
  magnétique},\ }\href {https://doi.org/10.1051/jphystap:018940030039300}
  {\bibfield  {journal} {\bibinfo  {journal} {J. Phys. Theor. Appl.}\ }\textbf
  {\bibinfo {volume} {3}},\ \bibinfo {pages} {393} (\bibinfo {year}
  {1894})}\BibitemShut {NoStop}%
\bibitem [{\citenamefont {Reimann}(2002)}]{Reimann2002}%
  \BibitemOpen
  \bibfield  {author} {\bibinfo {author} {\bibfnamefont {P.}~\bibnamefont
  {Reimann}},\ }\bibfield  {title} {\bibinfo {title} {Brownian motors: noisy
  transport far from equilibrium},\ }\href
  {https://doi.org/https://doi.org/10.1016/S0370-1573(01)00081-3} {\bibfield
  {journal} {\bibinfo  {journal} {Phys. Rep.}\ }\textbf {\bibinfo {volume}
  {361}},\ \bibinfo {pages} {57} (\bibinfo {year} {2002})}\BibitemShut
  {NoStop}%
\bibitem [{\citenamefont {Reichhardt}\ and\ \citenamefont
  {Reichhardt}(2017)}]{Reichhardt2017}%
  \BibitemOpen
  \bibfield  {author} {\bibinfo {author} {\bibfnamefont {C.~O.}\ \bibnamefont
  {Reichhardt}}\ and\ \bibinfo {author} {\bibfnamefont {C.}~\bibnamefont
  {Reichhardt}},\ }\bibfield  {title} {\bibinfo {title} {Ratchet effects in
  active matter systems},\ }\href
  {https://doi.org/https://doi.org/10.1146/annurev-conmatphys-031016-025522}
  {\bibfield  {journal} {\bibinfo  {journal} {Annu. Rev. Condens. Matter
  Phys.}\ }\textbf {\bibinfo {volume} {8}},\ \bibinfo {pages} {51} (\bibinfo
  {year} {2017})}\BibitemShut {NoStop}%
\bibitem [{\citenamefont {Borsley}\ \emph {et~al.}(2024)\citenamefont
  {Borsley}, \citenamefont {Leigh},\ and\ \citenamefont {Roberts}}]{Leigh2024}%
  \BibitemOpen
  \bibfield  {author} {\bibinfo {author} {\bibfnamefont {S.}~\bibnamefont
  {Borsley}}, \bibinfo {author} {\bibfnamefont {D.~A.}\ \bibnamefont {Leigh}},\
  and\ \bibinfo {author} {\bibfnamefont {B.~M.~W.}\ \bibnamefont {Roberts}},\
  }\bibfield  {title} {\bibinfo {title} {Molecular ratchets and kinetic
  asymmetry: Giving chemistry direction},\ }\href
  {https://doi.org/https://doi.org/10.1002/anie.202400495} {\bibfield
  {journal} {\bibinfo  {journal} {Angew. Chem. Int. Ed.}\ }\textbf {\bibinfo
  {volume} {63}},\ \bibinfo {pages} {e202400495} (\bibinfo {year}
  {2024})}\BibitemShut {NoStop}%
\bibitem [{\citenamefont {Aranson}\ and\ \citenamefont
  {Kramer}(2002)}]{Aranson2002}%
  \BibitemOpen
  \bibfield  {author} {\bibinfo {author} {\bibfnamefont {I.~S.}\ \bibnamefont
  {Aranson}}\ and\ \bibinfo {author} {\bibfnamefont {L.}~\bibnamefont
  {Kramer}},\ }\bibfield  {title} {\bibinfo {title} {The world of the complex
  ginzburg-landau equation},\ }\href {https://doi.org/10.1103/RevModPhys.74.99}
  {\bibfield  {journal} {\bibinfo  {journal} {Rev. Mod. Phys.}\ }\textbf
  {\bibinfo {volume} {74}},\ \bibinfo {pages} {99} (\bibinfo {year}
  {2002})}\BibitemShut {NoStop}%
\bibitem [{\citenamefont {Acebr\'on}\ \emph {et~al.}(2005)\citenamefont
  {Acebr\'on}, \citenamefont {Bonilla}, \citenamefont {P\'erez~Vicente},
  \citenamefont {Ritort},\ and\ \citenamefont {Spigler}}]{Ritort2005}%
  \BibitemOpen
  \bibfield  {author} {\bibinfo {author} {\bibfnamefont {J.~A.}\ \bibnamefont
  {Acebr\'on}}, \bibinfo {author} {\bibfnamefont {L.~L.}\ \bibnamefont
  {Bonilla}}, \bibinfo {author} {\bibfnamefont {C.~J.}\ \bibnamefont
  {P\'erez~Vicente}}, \bibinfo {author} {\bibfnamefont {F.}~\bibnamefont
  {Ritort}},\ and\ \bibinfo {author} {\bibfnamefont {R.}~\bibnamefont
  {Spigler}},\ }\bibfield  {title} {\bibinfo {title} {The kuramoto model: A
  simple paradigm for synchronization phenomena},\ }\href
  {https://doi.org/10.1103/RevModPhys.77.137} {\bibfield  {journal} {\bibinfo
  {journal} {Rev. Mod. Phys.}\ }\textbf {\bibinfo {volume} {77}},\ \bibinfo
  {pages} {137} (\bibinfo {year} {2005})}\BibitemShut {NoStop}%
\bibitem [{\citenamefont {Rouzaire}\ \emph
  {et~al.}(2025{\natexlab{b}})\citenamefont {Rouzaire}, \citenamefont
  {Rahmani}, \citenamefont {Pagonabarraga}, \citenamefont {Peruani},\ and\
  \citenamefont {Levis}}]{rouzaire2025_2}%
  \BibitemOpen
  \bibfield  {author} {\bibinfo {author} {\bibfnamefont {Y.}~\bibnamefont
  {Rouzaire}}, \bibinfo {author} {\bibfnamefont {P.}~\bibnamefont {Rahmani}},
  \bibinfo {author} {\bibfnamefont {I.}~\bibnamefont {Pagonabarraga}}, \bibinfo
  {author} {\bibfnamefont {F.}~\bibnamefont {Peruani}},\ and\ \bibinfo {author}
  {\bibfnamefont {D.}~\bibnamefont {Levis}},\ }\bibfield  {title} {\bibinfo
  {title} {Activity leads to topological phase transition in 2d populations of
  heterogeneous oscillators},\ }\href
  {https://doi.org/10.1103/PhysRevLett.134.188301} {\bibfield  {journal}
  {\bibinfo  {journal} {Phys. Rev. Lett.}\ }\textbf {\bibinfo {volume} {134}},\
  \bibinfo {pages} {188301} (\bibinfo {year} {2025}{\natexlab{b}})}\BibitemShut
  {NoStop}%
\bibitem [{\citenamefont {Dean}(1996)}]{dean}%
  \BibitemOpen
  \bibfield  {author} {\bibinfo {author} {\bibfnamefont {D.~S.}\ \bibnamefont
  {Dean}},\ }\bibfield  {title} {\bibinfo {title} {Langevin equation for the
  density of a system of interacting langevin processes},\ }\href
  {https://doi.org/10.1088/0305-4470/29/24/001} {\bibfield  {journal} {\bibinfo
   {journal} {J. Phys. A : Math. Gen.}\ }\textbf {\bibinfo {volume} {29}},\
  \bibinfo {pages} {L613–L617} (\bibinfo {year} {1996})}\BibitemShut
  {NoStop}%
\bibitem [{\citenamefont {Fodor}\ and\ \citenamefont
  {Marchetti}(2018)}]{Marchetti2018}%
  \BibitemOpen
  \bibfield  {author} {\bibinfo {author} {\bibfnamefont {E.}~\bibnamefont
  {Fodor}}\ and\ \bibinfo {author} {\bibfnamefont {M.~C.}\ \bibnamefont
  {Marchetti}},\ }\bibfield  {title} {\bibinfo {title} {The statistical physics
  of active matter: From self-catalytic colloids to living cells},\ }\href
  {https://doi.org/https://doi.org/10.1016/j.physa.2017.12.137} {\bibfield
  {journal} {\bibinfo  {journal} {Physica A}\ }\textbf {\bibinfo {volume}
  {504}},\ \bibinfo {pages} {106} (\bibinfo {year} {2018})}\BibitemShut
  {NoStop}%
\bibitem [{\citenamefont {Cates}\ and\ \citenamefont
  {Tailleur}(2015)}]{Cates2015}%
  \BibitemOpen
  \bibfield  {author} {\bibinfo {author} {\bibfnamefont {M.~E.}\ \bibnamefont
  {Cates}}\ and\ \bibinfo {author} {\bibfnamefont {J.}~\bibnamefont
  {Tailleur}},\ }\bibfield  {title} {\bibinfo {title} {Motility-induced phase
  separation},\ }\href
  {https://doi.org/10.1146/annurev-conmatphys-031214-014710} {\bibfield
  {journal} {\bibinfo  {journal} {Annu. Rev. Condens. Matter Phys.}\ }\textbf
  {\bibinfo {volume} {6}},\ \bibinfo {pages} {219} (\bibinfo {year}
  {2015})}\BibitemShut {NoStop}%
\bibitem [{\citenamefont {Chat\'e}(2020)}]{Chate2020}%
  \BibitemOpen
  \bibfield  {author} {\bibinfo {author} {\bibfnamefont {H.}~\bibnamefont
  {Chat\'e}},\ }\bibfield  {title} {\bibinfo {title} {Dry aligning dilute
  active matter},\ }\href
  {https://doi.org/10.1146/annurev-conmatphys-031119-050752} {\bibfield
  {journal} {\bibinfo  {journal} {Annu. Rev. Condens. Matter Phys.}\ }\textbf
  {\bibinfo {volume} {11}},\ \bibinfo {pages} {189} (\bibinfo {year}
  {2020})}\BibitemShut {NoStop}%
\bibitem [{\citenamefont {Archer}(2009)}]{Archer2009}%
  \BibitemOpen
  \bibfield  {author} {\bibinfo {author} {\bibfnamefont {A.~J.}\ \bibnamefont
  {Archer}},\ }\bibfield  {title} {\bibinfo {title} {Dynamical density
  functional theory for molecular and colloidal fluids: A microscopic approach
  to fluid mechanics},\ }\href {https://doi.org/10.1063/1.3054633} {\bibfield
  {journal} {\bibinfo  {journal} {J Chem. Phys.}\ }\textbf {\bibinfo {volume}
  {130}},\ \bibinfo {pages} {014509} (\bibinfo {year} {2009})}\BibitemShut
  {NoStop}%
\bibitem [{\citenamefont {Manacorda}\ and\ \citenamefont
  {Puglisi}(2017)}]{Manacorda2017}%
  \BibitemOpen
  \bibfield  {author} {\bibinfo {author} {\bibfnamefont {A.}~\bibnamefont
  {Manacorda}}\ and\ \bibinfo {author} {\bibfnamefont {A.}~\bibnamefont
  {Puglisi}},\ }\bibfield  {title} {\bibinfo {title} {Lattice model to derive
  the fluctuating hydrodynamics of active particles with inertia},\ }\href
  {https://doi.org/10.1103/PhysRevLett.119.208003} {\bibfield  {journal}
  {\bibinfo  {journal} {Phys. Rev. Lett.}\ }\textbf {\bibinfo {volume} {119}},\
  \bibinfo {pages} {208003} (\bibinfo {year} {2017})}\BibitemShut {NoStop}%
\bibitem [{\citenamefont {Manacorda}\ and\ \citenamefont
  {Fodor}(2025)}]{Manacorda2025}%
  \BibitemOpen
  \bibfield  {author} {\bibinfo {author} {\bibfnamefont {A.}~\bibnamefont
  {Manacorda}}\ and\ \bibinfo {author} {\bibfnamefont {E.}~\bibnamefont
  {Fodor}},\ }\bibfield  {title} {\bibinfo {title} {Diffusive oscillators
  capture the pulsating states of deformable particles},\ }\href
  {https://doi.org/10.1103/PhysRevE.111.L053401} {\bibfield  {journal}
  {\bibinfo  {journal} {Phys. Rev. E}\ }\textbf {\bibinfo {volume} {111}},\
  \bibinfo {pages} {L053401} (\bibinfo {year} {2025})}\BibitemShut {NoStop}%
\bibitem [{\citenamefont {Zhang}\ \emph {et~al.}(2025)\citenamefont {Zhang},
  \citenamefont {Manacorda},\ and\ \citenamefont {Fodor}}]{Zhang2025}%
  \BibitemOpen
  \bibfield  {author} {\bibinfo {author} {\bibfnamefont {Y.}~\bibnamefont
  {Zhang}}, \bibinfo {author} {\bibfnamefont {A.}~\bibnamefont {Manacorda}},\
  and\ \bibinfo {author} {\bibfnamefont {{\'E}.}~\bibnamefont {Fodor}},\
  }\bibfield  {title} {\bibinfo {title} {Species interconversion of deformable
  particles yields transient phase separation},\ }\href
  {https://doi.org/10.1088/1367-2630/adccf1} {\bibfield  {journal} {\bibinfo
  {journal} {New J. Phys}\ }\textbf {\bibinfo {volume} {27}},\ \bibinfo {pages}
  {043023} (\bibinfo {year} {2025})}\BibitemShut {NoStop}%
\bibitem [{\citenamefont {Antonsen}\ \emph {et~al.}(2008)\citenamefont
  {Antonsen}, \citenamefont {Faghih}, \citenamefont {Girvan}, \citenamefont
  {Ott},\ and\ \citenamefont {Platig}}]{Antonsen2008}%
  \BibitemOpen
  \bibfield  {author} {\bibinfo {author} {\bibfnamefont {J.}~\bibnamefont
  {Antonsen}, \bibfnamefont {T.~M.}}, \bibinfo {author} {\bibfnamefont {R.~T.}\
  \bibnamefont {Faghih}}, \bibinfo {author} {\bibfnamefont {M.}~\bibnamefont
  {Girvan}}, \bibinfo {author} {\bibfnamefont {E.}~\bibnamefont {Ott}},\ and\
  \bibinfo {author} {\bibfnamefont {J.}~\bibnamefont {Platig}},\ }\bibfield
  {title} {\bibinfo {title} {External periodic driving of large systems of
  globally coupled phase oscillators},\ }\href
  {https://doi.org/10.1063/1.2952447} {\bibfield  {journal} {\bibinfo
  {journal} {Chaos}\ }\textbf {\bibinfo {volume} {18}},\ \bibinfo {pages}
  {037112} (\bibinfo {year} {2008})}\BibitemShut {NoStop}%
\bibitem [{\citenamefont {Childs}\ and\ \citenamefont
  {Strogatz}(2008)}]{Strogatz2008}%
  \BibitemOpen
  \bibfield  {author} {\bibinfo {author} {\bibfnamefont {L.~M.}\ \bibnamefont
  {Childs}}\ and\ \bibinfo {author} {\bibfnamefont {S.~H.}\ \bibnamefont
  {Strogatz}},\ }\bibfield  {title} {\bibinfo {title} {Stability diagram for
  the forced kuramoto model},\ }\href {https://doi.org/10.1063/1.3049136}
  {\bibfield  {journal} {\bibinfo  {journal} {Chaos}\ }\textbf {\bibinfo
  {volume} {18}},\ \bibinfo {pages} {043128} (\bibinfo {year}
  {2008})}\BibitemShut {NoStop}%
\bibitem [{\citenamefont {Cestnik}\ and\ \citenamefont
  {Pikovsky}(2022)}]{cestnik2022}%
  \BibitemOpen
  \bibfield  {author} {\bibinfo {author} {\bibfnamefont {R.}~\bibnamefont
  {Cestnik}}\ and\ \bibinfo {author} {\bibfnamefont {A.}~\bibnamefont
  {Pikovsky}},\ }\bibfield  {title} {\bibinfo {title} {Hierarchy of exact
  low-dimensional reductions for populations of coupled oscillators},\ }\href
  {https://doi.org/10.1103/PhysRevLett.128.054101} {\bibfield  {journal}
  {\bibinfo  {journal} {Phys. Rev. Lett.}\ }\textbf {\bibinfo {volume} {128}},\
  \bibinfo {pages} {054101} (\bibinfo {year} {2022})}\BibitemShut {NoStop}%
\bibitem [{\citenamefont {Buend\'{\i}a}(2025)}]{Buendia2025}%
  \BibitemOpen
  \bibfield  {author} {\bibinfo {author} {\bibfnamefont {V.}~\bibnamefont
  {Buend\'{\i}a}},\ }\bibfield  {title} {\bibinfo {title} {Mesoscopic theory
  for coupled stochastic oscillators},\ }\href
  {https://doi.org/10.1103/PhysRevLett.134.197201} {\bibfield  {journal}
  {\bibinfo  {journal} {Phys. Rev. Lett.}\ }\textbf {\bibinfo {volume} {134}},\
  \bibinfo {pages} {197201} (\bibinfo {year} {2025})}\BibitemShut {NoStop}%
\bibitem [{\citenamefont {Majumder}\ \emph {et~al.}(2025)\citenamefont
  {Majumder}, \citenamefont {Barré},\ and\ \citenamefont {Gupta}}]{gupta2025}%
  \BibitemOpen
  \bibfield  {author} {\bibinfo {author} {\bibfnamefont {R.}~\bibnamefont
  {Majumder}}, \bibinfo {author} {\bibfnamefont {J.}~\bibnamefont {Barré}},\
  and\ \bibinfo {author} {\bibfnamefont {S.}~\bibnamefont {Gupta}},\ }\href
  {https://arxiv.org/abs/2510.02448} {\bibinfo {title} {Finite-size
  fluctuations for stochastic coupled oscillators: A general theory}} (\bibinfo
  {year} {2025}),\ \Eprint {https://arxiv.org/abs/2510.02448} {arXiv:2510.02448
  [cond-mat.stat-mech]} \BibitemShut {NoStop}%
\bibitem [{\citenamefont {Chatzittofi}\ \emph {et~al.}(2023)\citenamefont
  {Chatzittofi}, \citenamefont {Golestanian},\ and\ \citenamefont
  {Agudo-Canalejo}}]{chatzittofi2023}%
  \BibitemOpen
  \bibfield  {author} {\bibinfo {author} {\bibfnamefont {M.}~\bibnamefont
  {Chatzittofi}}, \bibinfo {author} {\bibfnamefont {R.}~\bibnamefont
  {Golestanian}},\ and\ \bibinfo {author} {\bibfnamefont {J.}~\bibnamefont
  {Agudo-Canalejo}},\ }\bibfield  {title} {\bibinfo {title} {Collective
  synchronization of dissipatively-coupled noise-activated processes},\ }\href
  {https://doi.org/10.1088/1367-2630/acf2bc} {\bibfield  {journal} {\bibinfo
  {journal} {New J. Phys.}\ }\textbf {\bibinfo {volume} {25}},\ \bibinfo
  {pages} {093014} (\bibinfo {year} {2023})}\BibitemShut {NoStop}%
\bibitem [{\citenamefont {Martin}\ \emph {et~al.}(2025)\citenamefont {Martin},
  \citenamefont {Seara}, \citenamefont {Avni}, \citenamefont {Fruchart},\ and\
  \citenamefont {Vitelli}}]{martin2025}%
  \BibitemOpen
  \bibfield  {author} {\bibinfo {author} {\bibfnamefont {D.}~\bibnamefont
  {Martin}}, \bibinfo {author} {\bibfnamefont {D.}~\bibnamefont {Seara}},
  \bibinfo {author} {\bibfnamefont {Y.}~\bibnamefont {Avni}}, \bibinfo {author}
  {\bibfnamefont {M.}~\bibnamefont {Fruchart}},\ and\ \bibinfo {author}
  {\bibfnamefont {V.}~\bibnamefont {Vitelli}},\ }\bibfield  {title} {\bibinfo
  {title} {Transition to collective motion in nonreciprocal active matter:
  Coarse graining agent-based models into fluctuating hydrodynamics},\ }\href
  {https://doi.org/10.1103/PhysRevX.15.041015} {\bibfield  {journal} {\bibinfo
  {journal} {Phys. Rev. X}\ }\textbf {\bibinfo {volume} {15}},\ \bibinfo
  {pages} {041015} (\bibinfo {year} {2025})}\BibitemShut {NoStop}%
\bibitem [{\citenamefont {Blom}\ \emph {et~al.}(2025)\citenamefont {Blom},
  \citenamefont {Thiele},\ and\ \citenamefont {Godec}}]{Blom2025}%
  \BibitemOpen
  \bibfield  {author} {\bibinfo {author} {\bibfnamefont {K.}~\bibnamefont
  {Blom}}, \bibinfo {author} {\bibfnamefont {U.}~\bibnamefont {Thiele}},\ and\
  \bibinfo {author} {\bibfnamefont {A.~c.~v.}\ \bibnamefont {Godec}},\
  }\bibfield  {title} {\bibinfo {title} {Local order controls the onset of
  oscillations in the nonreciprocal ising model},\ }\href
  {https://doi.org/10.1103/PhysRevE.111.024207} {\bibfield  {journal} {\bibinfo
   {journal} {Phys. Rev. E}\ }\textbf {\bibinfo {volume} {111}},\ \bibinfo
  {pages} {024207} (\bibinfo {year} {2025})}\BibitemShut {NoStop}%
\bibitem [{\citenamefont {Fodor}\ \emph {et~al.}(2016)\citenamefont {Fodor},
  \citenamefont {Nardini}, \citenamefont {Cates}, \citenamefont {Tailleur},
  \citenamefont {Visco},\ and\ \citenamefont {van Wijland}}]{Nardini2016}%
  \BibitemOpen
  \bibfield  {author} {\bibinfo {author} {\bibfnamefont {E.}~\bibnamefont
  {Fodor}}, \bibinfo {author} {\bibfnamefont {C.}~\bibnamefont {Nardini}},
  \bibinfo {author} {\bibfnamefont {M.~E.}\ \bibnamefont {Cates}}, \bibinfo
  {author} {\bibfnamefont {J.}~\bibnamefont {Tailleur}}, \bibinfo {author}
  {\bibfnamefont {P.}~\bibnamefont {Visco}},\ and\ \bibinfo {author}
  {\bibfnamefont {F.}~\bibnamefont {van Wijland}},\ }\bibfield  {title}
  {\bibinfo {title} {How far from equilibrium is active matter?},\ }\href
  {https://doi.org/10.1103/PhysRevLett.117.038103} {\bibfield  {journal}
  {\bibinfo  {journal} {Phys. Rev. Lett.}\ }\textbf {\bibinfo {volume} {117}},\
  \bibinfo {pages} {038103} (\bibinfo {year} {2016})}\BibitemShut {NoStop}%
\bibitem [{\citenamefont {Nardini}\ \emph {et~al.}(2017)\citenamefont
  {Nardini}, \citenamefont {Fodor}, \citenamefont {Tjhung}, \citenamefont {van
  Wijland}, \citenamefont {Tailleur},\ and\ \citenamefont
  {Cates}}]{Nardini2017}%
  \BibitemOpen
  \bibfield  {author} {\bibinfo {author} {\bibfnamefont {C.}~\bibnamefont
  {Nardini}}, \bibinfo {author} {\bibfnamefont {E.}~\bibnamefont {Fodor}},
  \bibinfo {author} {\bibfnamefont {E.}~\bibnamefont {Tjhung}}, \bibinfo
  {author} {\bibfnamefont {F.}~\bibnamefont {van Wijland}}, \bibinfo {author}
  {\bibfnamefont {J.}~\bibnamefont {Tailleur}},\ and\ \bibinfo {author}
  {\bibfnamefont {M.~E.}\ \bibnamefont {Cates}},\ }\bibfield  {title} {\bibinfo
  {title} {Entropy production in field theories without time-reversal symmetry:
  Quantifying the non-equilibrium character of active matter},\ }\href
  {https://doi.org/10.1103/PhysRevX.7.021007} {\bibfield  {journal} {\bibinfo
  {journal} {Phys. Rev. X}\ }\textbf {\bibinfo {volume} {7}},\ \bibinfo {pages}
  {021007} (\bibinfo {year} {2017})}\BibitemShut {NoStop}%
\bibitem [{\citenamefont {Fodor}\ \emph {et~al.}(2022)\citenamefont {Fodor},
  \citenamefont {Jack},\ and\ \citenamefont {Cates}}]{Jack2022}%
  \BibitemOpen
  \bibfield  {author} {\bibinfo {author} {\bibfnamefont {{\'E}.}~\bibnamefont
  {Fodor}}, \bibinfo {author} {\bibfnamefont {R.~L.}\ \bibnamefont {Jack}},\
  and\ \bibinfo {author} {\bibfnamefont {M.~E.}\ \bibnamefont {Cates}},\
  }\bibfield  {title} {\bibinfo {title} {Irreversibility and biased ensembles
  in active matter: Insights from stochastic thermodynamics},\ }\href
  {https://doi.org/10.1146/annurev-conmatphys-031720-032419} {\bibfield
  {journal} {\bibinfo  {journal} {Annu. Rev. Condens. Matter Phys.}\ }\textbf
  {\bibinfo {volume} {13}},\ \bibinfo {pages} {215} (\bibinfo {year}
  {2022})}\BibitemShut {NoStop}%
\bibitem [{\citenamefont {Liebchen}\ and\ \citenamefont
  {Levis}(2017)}]{liebchen2017}%
  \BibitemOpen
  \bibfield  {author} {\bibinfo {author} {\bibfnamefont {B.}~\bibnamefont
  {Liebchen}}\ and\ \bibinfo {author} {\bibfnamefont {D.}~\bibnamefont
  {Levis}},\ }\bibfield  {title} {\bibinfo {title} {Collective behavior of
  chiral active matter: Pattern formation and enhanced flocking},\ }\href
  {https://doi.org/10.1103/PhysRevLett.119.058002} {\bibfield  {journal}
  {\bibinfo  {journal} {Phys. Rev. Lett.}\ }\textbf {\bibinfo {volume} {119}},\
  \bibinfo {pages} {058002} (\bibinfo {year} {2017})}\BibitemShut {NoStop}%
\bibitem [{\citenamefont {Aranson}\ \emph {et~al.}(1998)\citenamefont
  {Aranson}, \citenamefont {Chat\'e},\ and\ \citenamefont
  {Tang}}]{aranson1998}%
  \BibitemOpen
  \bibfield  {author} {\bibinfo {author} {\bibfnamefont {I.~S.}\ \bibnamefont
  {Aranson}}, \bibinfo {author} {\bibfnamefont {H.}~\bibnamefont {Chat\'e}},\
  and\ \bibinfo {author} {\bibfnamefont {L.-H.}\ \bibnamefont {Tang}},\
  }\bibfield  {title} {\bibinfo {title} {Spiral motion in a noisy complex
  ginzburg-landau equation},\ }\href
  {https://doi.org/10.1103/PhysRevLett.80.2646} {\bibfield  {journal} {\bibinfo
   {journal} {Phys. Rev. Lett.}\ }\textbf {\bibinfo {volume} {80}},\ \bibinfo
  {pages} {2646} (\bibinfo {year} {1998})}\BibitemShut {NoStop}%
\bibitem [{\citenamefont {Brito}\ \emph {et~al.}(2003)\citenamefont {Brito},
  \citenamefont {Aranson},\ and\ \citenamefont {Chat\'e}}]{brito2003}%
  \BibitemOpen
  \bibfield  {author} {\bibinfo {author} {\bibfnamefont {C.}~\bibnamefont
  {Brito}}, \bibinfo {author} {\bibfnamefont {I.~S.}\ \bibnamefont {Aranson}},\
  and\ \bibinfo {author} {\bibfnamefont {H.}~\bibnamefont {Chat\'e}},\
  }\bibfield  {title} {\bibinfo {title} {Vortex glass and vortex liquid in
  oscillatory media},\ }\href {https://doi.org/10.1103/PhysRevLett.90.068301}
  {\bibfield  {journal} {\bibinfo  {journal} {Phys. Rev. Lett.}\ }\textbf
  {\bibinfo {volume} {90}},\ \bibinfo {pages} {068301} (\bibinfo {year}
  {2003})}\BibitemShut {NoStop}%
\bibitem [{\citenamefont {Fruchart}\ and\ \citenamefont
  {Vitelli}(2026)}]{fruchart2026}%
  \BibitemOpen
  \bibfield  {author} {\bibinfo {author} {\bibfnamefont {M.}~\bibnamefont
  {Fruchart}}\ and\ \bibinfo {author} {\bibfnamefont {V.}~\bibnamefont
  {Vitelli}},\ }\href {https://arxiv.org/abs/2602.11111} {\bibinfo {title}
  {Nonreciprocal many-body physics}} (\bibinfo {year} {2026}),\ \Eprint
  {https://arxiv.org/abs/2602.11111} {arXiv:2602.11111 [cond-mat.stat-mech]}
  \BibitemShut {NoStop}%
\bibitem [{\citenamefont {Shankar}\ \emph
  {et~al.}(2022{\natexlab{b}})\citenamefont {Shankar}, \citenamefont {Raju},\
  and\ \citenamefont {Mahadevan}}]{shankar2022drops}%
  \BibitemOpen
  \bibfield  {author} {\bibinfo {author} {\bibfnamefont {S.}~\bibnamefont
  {Shankar}}, \bibinfo {author} {\bibfnamefont {V.}~\bibnamefont {Raju}},\ and\
  \bibinfo {author} {\bibfnamefont {L.}~\bibnamefont {Mahadevan}},\ }\bibfield
  {title} {\bibinfo {title} {Optimal transport and control of active drops},\
  }\href {https://doi.org/10.1073/pnas.2121985119} {\bibfield  {journal}
  {\bibinfo  {journal} {Proc. Natl. Acad. Sci. USA}\ }\textbf {\bibinfo
  {volume} {119}},\ \bibinfo {pages} {e2121985119} (\bibinfo {year}
  {2022}{\natexlab{b}})}\BibitemShut {NoStop}%
\bibitem [{\citenamefont {Davis}\ \emph {et~al.}(2024)\citenamefont {Davis},
  \citenamefont {Proesmans},\ and\ \citenamefont {Fodor}}]{Davis2025}%
  \BibitemOpen
  \bibfield  {author} {\bibinfo {author} {\bibfnamefont {L.~K.}\ \bibnamefont
  {Davis}}, \bibinfo {author} {\bibfnamefont {K.}~\bibnamefont {Proesmans}},\
  and\ \bibinfo {author} {\bibfnamefont {E.}~\bibnamefont {Fodor}},\ }\bibfield
   {title} {\bibinfo {title} {Active matter under control: Insights from
  response theory},\ }\href {https://doi.org/10.1103/PhysRevX.14.011012}
  {\bibfield  {journal} {\bibinfo  {journal} {Phys. Rev. X}\ }\textbf {\bibinfo
  {volume} {14}},\ \bibinfo {pages} {011012} (\bibinfo {year}
  {2024})}\BibitemShut {NoStop}%
\bibitem [{\citenamefont {Garcia-Millan}\ \emph {et~al.}(2025)\citenamefont
  {Garcia-Millan}, \citenamefont {Sch\"uttler}, \citenamefont {Cates},\ and\
  \citenamefont {Loos}}]{fbgp-qpvv}%
  \BibitemOpen
  \bibfield  {author} {\bibinfo {author} {\bibfnamefont {R.}~\bibnamefont
  {Garcia-Millan}}, \bibinfo {author} {\bibfnamefont {J.}~\bibnamefont
  {Sch\"uttler}}, \bibinfo {author} {\bibfnamefont {M.~E.}\ \bibnamefont
  {Cates}},\ and\ \bibinfo {author} {\bibfnamefont {S.~A.~M.}\ \bibnamefont
  {Loos}},\ }\bibfield  {title} {\bibinfo {title} {Optimal closed-loop control
  of active particles and a minimal information engine},\ }\href
  {https://doi.org/10.1103/fbgp-qpvv} {\bibfield  {journal} {\bibinfo
  {journal} {Phys. Rev. Lett.}\ }\textbf {\bibinfo {volume} {135}},\ \bibinfo
  {pages} {088301} (\bibinfo {year} {2025})}\BibitemShut {NoStop}%
\bibitem [{\citenamefont {Soriani}\ \emph {et~al.}(2025)\citenamefont
  {Soriani}, \citenamefont {Tjhung}, \citenamefont {Fodor},\ and\ \citenamefont
  {Markovich}}]{soriani2025}%
  \BibitemOpen
  \bibfield  {author} {\bibinfo {author} {\bibfnamefont {A.}~\bibnamefont
  {Soriani}}, \bibinfo {author} {\bibfnamefont {E.}~\bibnamefont {Tjhung}},
  \bibinfo {author} {\bibfnamefont {{\'E}.}~\bibnamefont {Fodor}},\ and\
  \bibinfo {author} {\bibfnamefont {T.}~\bibnamefont {Markovich}},\ }\href
  {https://arxiv.org/abs/2504.19285} {\bibinfo {title} {Control of active field
  theories at minimal dissipation}} (\bibinfo {year} {2025}),\ \Eprint
  {https://arxiv.org/abs/2504.19285} {arXiv:2504.19285 [cond-mat.stat-mech]}
  \BibitemShut {NoStop}%
\bibitem [{\citenamefont {Alvarado}\ \emph {et~al.}(2026)\citenamefont
  {Alvarado}, \citenamefont {Teich}, \citenamefont {Sivak},\ and\ \citenamefont
  {Bechhoefer}}]{alvarado2025}%
  \BibitemOpen
  \bibfield  {author} {\bibinfo {author} {\bibfnamefont {J.}~\bibnamefont
  {Alvarado}}, \bibinfo {author} {\bibfnamefont {E.~G.}\ \bibnamefont {Teich}},
  \bibinfo {author} {\bibfnamefont {D.~A.}\ \bibnamefont {Sivak}},\ and\
  \bibinfo {author} {\bibfnamefont {J.}~\bibnamefont {Bechhoefer}},\ }\bibfield
   {title} {\bibinfo {title} {Optimal control in soft and active matter},\
  }\href
  {https://doi.org/https://doi.org/10.1146/annurev-conmatphys-031324-031350}
  {\bibfield  {journal} {\bibinfo  {journal} {Annu. Rev. Condens. Matter
  Phys.}\ }\textbf {\bibinfo {volume} {17}},\ \bibinfo {pages} {327} (\bibinfo
  {year} {2026})}\BibitemShut {NoStop}%
\bibitem [{\citenamefont {Krishnan}\ \emph {et~al.}(2026)\citenamefont
  {Krishnan}, \citenamefont {Sinha},\ and\ \citenamefont
  {Mahadevan}}]{krishnan2024}%
  \BibitemOpen
  \bibfield  {author} {\bibinfo {author} {\bibfnamefont {V.}~\bibnamefont
  {Krishnan}}, \bibinfo {author} {\bibfnamefont {S.}~\bibnamefont {Sinha}},\
  and\ \bibinfo {author} {\bibfnamefont {L.}~\bibnamefont {Mahadevan}},\
  }\bibfield  {title} {\bibinfo {title} {Hamiltonian bridge for scale-dependent
  optimal control of particles and fields},\ }\href
  {https://doi.org/https://doi.org/10.1016/j.newton.2025.100365} {\bibfield
  {journal} {\bibinfo  {journal} {Newton}\ }\textbf {\bibinfo {volume} {2}},\
  \bibinfo {pages} {100365} (\bibinfo {year} {2026})}\BibitemShut {NoStop}%
\bibitem [{\citenamefont {Norton}\ \emph {et~al.}(2020)\citenamefont {Norton},
  \citenamefont {Grover}, \citenamefont {Hagan},\ and\ \citenamefont
  {Fraden}}]{Norton2020}%
  \BibitemOpen
  \bibfield  {author} {\bibinfo {author} {\bibfnamefont {M.~M.}\ \bibnamefont
  {Norton}}, \bibinfo {author} {\bibfnamefont {P.}~\bibnamefont {Grover}},
  \bibinfo {author} {\bibfnamefont {M.~F.}\ \bibnamefont {Hagan}},\ and\
  \bibinfo {author} {\bibfnamefont {S.}~\bibnamefont {Fraden}},\ }\bibfield
  {title} {\bibinfo {title} {Optimal control of active nematics},\ }\href
  {https://doi.org/10.1103/PhysRevLett.125.178005} {\bibfield  {journal}
  {\bibinfo  {journal} {Phys. Rev. Lett.}\ }\textbf {\bibinfo {volume} {125}},\
  \bibinfo {pages} {178005} (\bibinfo {year} {2020})}\BibitemShut {NoStop}%
\bibitem [{\citenamefont {Shankar}\ \emph {et~al.}(2024)\citenamefont
  {Shankar}, \citenamefont {Scharrer}, \citenamefont {Bowick},\ and\
  \citenamefont {Marchetti}}]{Shankar2024}%
  \BibitemOpen
  \bibfield  {author} {\bibinfo {author} {\bibfnamefont {S.}~\bibnamefont
  {Shankar}}, \bibinfo {author} {\bibfnamefont {L.~V.~D.}\ \bibnamefont
  {Scharrer}}, \bibinfo {author} {\bibfnamefont {M.~J.}\ \bibnamefont
  {Bowick}},\ and\ \bibinfo {author} {\bibfnamefont {M.~C.}\ \bibnamefont
  {Marchetti}},\ }\bibfield  {title} {\bibinfo {title} {Design rules for
  controlling active topological defects},\ }\href
  {https://doi.org/10.1073/pnas.2400933121} {\bibfield  {journal} {\bibinfo
  {journal} {Proc. Natl. Acad. Sci. USA}\ }\textbf {\bibinfo {volume} {121}},\
  \bibinfo {pages} {e2400933121} (\bibinfo {year} {2024})}\BibitemShut
  {NoStop}%
\bibitem [{\citenamefont {Ghosh}\ \emph {et~al.}(2024)\citenamefont {Ghosh},
  \citenamefont {Joshi}, \citenamefont {Baskaran},\ and\ \citenamefont
  {Hagan}}]{ghosh2024}%
  \BibitemOpen
  \bibfield  {author} {\bibinfo {author} {\bibfnamefont {S.}~\bibnamefont
  {Ghosh}}, \bibinfo {author} {\bibfnamefont {C.}~\bibnamefont {Joshi}},
  \bibinfo {author} {\bibfnamefont {A.}~\bibnamefont {Baskaran}},\ and\
  \bibinfo {author} {\bibfnamefont {M.~F.}\ \bibnamefont {Hagan}},\ }\bibfield
  {title} {\bibinfo {title} {Spatiotemporal control of structure and dynamics
  in a polar active fluid},\ }\href {https://doi.org/10.1039/D4SM00547C}
  {\bibfield  {journal} {\bibinfo  {journal} {Soft Matter}\ }\textbf {\bibinfo
  {volume} {20}},\ \bibinfo {pages} {7059} (\bibinfo {year}
  {2024})}\BibitemShut {NoStop}%
\bibitem [{\citenamefont {Ghosh}\ \emph {et~al.}(2025)\citenamefont {Ghosh},
  \citenamefont {Baskaran},\ and\ \citenamefont {Hagan}}]{ghosh2025}%
  \BibitemOpen
  \bibfield  {author} {\bibinfo {author} {\bibfnamefont {S.}~\bibnamefont
  {Ghosh}}, \bibinfo {author} {\bibfnamefont {A.}~\bibnamefont {Baskaran}},\
  and\ \bibinfo {author} {\bibfnamefont {M.~F.}\ \bibnamefont {Hagan}},\
  }\bibfield  {title} {\bibinfo {title} {Achieving designed texture and flows
  in bulk active nematics using optimal control theory},\ }\href
  {https://doi.org/10.1063/5.0244046} {\bibfield  {journal} {\bibinfo
  {journal} {J. Chem. Phys.}\ }\textbf {\bibinfo {volume} {162}},\ \bibinfo
  {pages} {134902} (\bibinfo {year} {2025})}\BibitemShut {NoStop}%
\bibitem [{\citenamefont {Geerds}\ \emph {et~al.}(2025)\citenamefont {Geerds},
  \citenamefont {Singh}, \citenamefont {Dedenon}, \citenamefont {Pearce},
  \citenamefont {Jülicher}, \citenamefont {Sbalzarini},\ and\ \citenamefont
  {Kruse}}]{geerds2025}%
  \BibitemOpen
  \bibfield  {author} {\bibinfo {author} {\bibfnamefont {B.~C.}\ \bibnamefont
  {Geerds}}, \bibinfo {author} {\bibfnamefont {A.}~\bibnamefont {Singh}},
  \bibinfo {author} {\bibfnamefont {M.}~\bibnamefont {Dedenon}}, \bibinfo
  {author} {\bibfnamefont {D.~J.~G.}\ \bibnamefont {Pearce}}, \bibinfo {author}
  {\bibfnamefont {F.}~\bibnamefont {Jülicher}}, \bibinfo {author}
  {\bibfnamefont {I.~F.}\ \bibnamefont {Sbalzarini}},\ and\ \bibinfo {author}
  {\bibfnamefont {K.}~\bibnamefont {Kruse}},\ }\href
  {https://arxiv.org/abs/2511.21359} {\bibinfo {title} {Spatiotemporal control
  of charge +1 topological defects in polar active matter}} (\bibinfo {year}
  {2025}),\ \Eprint {https://arxiv.org/abs/2511.21359} {arXiv:2511.21359
  [cond-mat.soft]} \BibitemShut {NoStop}%
\bibitem [{\citenamefont {Skogvoll}\ \emph {et~al.}(2023)\citenamefont
  {Skogvoll}, \citenamefont {R{\o}nning}, \citenamefont {Salvalaglio},\ and\
  \citenamefont {Angheluta}}]{skogvoll2023}%
  \BibitemOpen
  \bibfield  {author} {\bibinfo {author} {\bibfnamefont {V.}~\bibnamefont
  {Skogvoll}}, \bibinfo {author} {\bibfnamefont {J.}~\bibnamefont
  {R{\o}nning}}, \bibinfo {author} {\bibfnamefont {M.}~\bibnamefont
  {Salvalaglio}},\ and\ \bibinfo {author} {\bibfnamefont {L.}~\bibnamefont
  {Angheluta}},\ }\bibfield  {title} {\bibinfo {title} {A unified field theory
  of topological defects and non-linear local excitations},\ }\href
  {https://doi.org/10.1038/s41524-023-01077-6} {\bibfield  {journal} {\bibinfo
  {journal} {npj Comput. Mater.}\ }\textbf {\bibinfo {volume} {9}},\ \bibinfo
  {pages} {122} (\bibinfo {year} {2023})}\BibitemShut {NoStop}%
\bibitem [{\citenamefont {Romano}\ \emph {et~al.}(2023)\citenamefont {Romano},
  \citenamefont {Mahault},\ and\ \citenamefont {Golestanian}}]{romano2023}%
  \BibitemOpen
  \bibfield  {author} {\bibinfo {author} {\bibfnamefont {J.}~\bibnamefont
  {Romano}}, \bibinfo {author} {\bibfnamefont {B.}~\bibnamefont {Mahault}},\
  and\ \bibinfo {author} {\bibfnamefont {R.}~\bibnamefont {Golestanian}},\
  }\bibfield  {title} {\bibinfo {title} {Dynamical theory of topological
  defects i: the multivalued solution of the diffusion equation},\ }\href
  {https://doi.org/10.1088/1742-5468/aceb57} {\bibfield  {journal} {\bibinfo
  {journal} {J. Stat. Mech.: Theory Exp.}\ }\textbf {\bibinfo {volume}
  {2023}},\ \bibinfo {pages} {083211}}\BibitemShut {NoStop}%
\bibitem [{\citenamefont {Romano}\ \emph {et~al.}(2024)\citenamefont {Romano},
  \citenamefont {Mahault},\ and\ \citenamefont {Golestanian}}]{romano2024}%
  \BibitemOpen
  \bibfield  {author} {\bibinfo {author} {\bibfnamefont {J.}~\bibnamefont
  {Romano}}, \bibinfo {author} {\bibfnamefont {B.}~\bibnamefont {Mahault}},\
  and\ \bibinfo {author} {\bibfnamefont {R.}~\bibnamefont {Golestanian}},\
  }\bibfield  {title} {\bibinfo {title} {Dynamical theory of topological
  defects ii: universal aspects of defect motion},\ }\href
  {https://doi.org/10.1088/1742-5468/ad2ddb} {\bibfield  {journal} {\bibinfo
  {journal} {J. Stat. Mech.: Theory Exp.}\ }\textbf {\bibinfo {volume}
  {2024}},\ \bibinfo {pages} {033208}}\BibitemShut {NoStop}%
\bibitem [{\citenamefont {Rana}\ and\ \citenamefont
  {Golestanian}(2024)}]{rana2024}%
  \BibitemOpen
  \bibfield  {author} {\bibinfo {author} {\bibfnamefont {N.}~\bibnamefont
  {Rana}}\ and\ \bibinfo {author} {\bibfnamefont {R.}~\bibnamefont
  {Golestanian}},\ }\bibfield  {title} {\bibinfo {title} {Defect solutions of
  the nonreciprocal cahn-hilliard model: Spirals and targets},\ }\href
  {https://doi.org/10.1103/PhysRevLett.133.078301} {\bibfield  {journal}
  {\bibinfo  {journal} {Phys. Rev. Lett.}\ }\textbf {\bibinfo {volume} {133}},\
  \bibinfo {pages} {078301} (\bibinfo {year} {2024})}\BibitemShut {NoStop}%
\bibitem [{\citenamefont {Munkres}(1957)}]{munkres1957}%
  \BibitemOpen
  \bibfield  {author} {\bibinfo {author} {\bibfnamefont {J.}~\bibnamefont
  {Munkres}},\ }\bibfield  {title} {\bibinfo {title} {Algorithms for the
  assignment and transportation problems},\ }\href
  {https://doi.org/10.1137/0105003} {\bibfield  {journal} {\bibinfo  {journal}
  {J. Soc. Indust. Appl. Math.}\ }\textbf {\bibinfo {volume} {5}},\ \bibinfo
  {pages} {32} (\bibinfo {year} {1957})}\BibitemShut {NoStop}%
\bibitem [{\citenamefont {Glasserman}(2003)}]{Glasserman}%
  \BibitemOpen
  \bibfield  {author} {\bibinfo {author} {\bibfnamefont {P.}~\bibnamefont
  {Glasserman}},\ }\href {https://doi.org/10.1007/978-0-387-21617-1} {\emph
  {\bibinfo {title} {Monte Carlo Methods in Financial Engineering}}}\ (\bibinfo
  {year} {2003})\BibitemShut {NoStop}%
\bibitem [{\citenamefont {Cai}\ \emph {et~al.}(2019)\citenamefont {Cai},
  \citenamefont {Chat\'e}, \citenamefont {Ma},\ and\ \citenamefont
  {Shi}}]{cai2019}%
  \BibitemOpen
  \bibfield  {author} {\bibinfo {author} {\bibfnamefont {L.-b.}\ \bibnamefont
  {Cai}}, \bibinfo {author} {\bibfnamefont {H.}~\bibnamefont {Chat\'e}},
  \bibinfo {author} {\bibfnamefont {Y.-q.}\ \bibnamefont {Ma}},\ and\ \bibinfo
  {author} {\bibfnamefont {X.-q.}\ \bibnamefont {Shi}},\ }\bibfield  {title}
  {\bibinfo {title} {Dynamical subclasses of dry active nematics},\ }\href
  {https://doi.org/10.1103/PhysRevE.99.010601} {\bibfield  {journal} {\bibinfo
  {journal} {Phys. Rev. E}\ }\textbf {\bibinfo {volume} {99}},\ \bibinfo
  {pages} {010601} (\bibinfo {year} {2019})}\BibitemShut {NoStop}%
\end{thebibliography}%


\clearpage
\onecolumngrid

\renewcommand{\theequation}{S\arabic{equation}}
\renewcommand{\thesection}{S\arabic{section}}
\renewcommand{\thetable}{S\arabic{table}}
\renewcommand{\thefigure}{S\arabic{figure}}
\setcounter{equation}{0}
\setcounter{section}{0}
\setcounter{figure}{0}
\setcounter{table}{0}

\begin{center}
{\large\bf Supplementary material: Topology of pulsating active matter}\\[10pt]
Luca Casagrande,$^{1}$ Alessandro Manacorda,$^{1,2}$ and \'Etienne Fodor$^{1}$\\[4pt]
\textit{$^{1}$Department of Physics and Materials Science, University of Luxembourg, L-1511 Luxembourg City, Luxembourg}\\
\textit{$^{2}$CNR Institute of Complex Systems, Uos Sapienza, Piazzale A. Moro 5, 00185 Rome, Italy}
\end{center}
\vspace{1em}

\section{NUMERICAL METHODS}
\label{sec:numerical}

\subsection{Particle-based simulations}
\label{subsec:particle}

\begin{figure}[b]
    \centering
     \includegraphics[width=0.5\columnwidth]{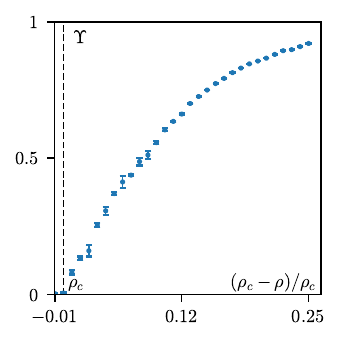}
    \caption{Mean pulsation rate $\Upsilon$ [Eq.~(S1)] as a function of the distance to the arrest transition $(\rho_c-\rho)/\rho_c$ in particle-based simulations [Eq.~(1)]. $\Upsilon$ vanishes at $\rho=\rho_c$ (dashed line), defining the critical density at which phase dynamics become fully arrested by repulsive interactions; $\Upsilon\to1$ represents cycling at low density. Error bars denote the standard error of the mean over $N_{\rm seeds} = 4$ independent simulation realizations at each density,
$\sigma_\Upsilon = {\rm std}(\{\Upsilon_i\}_{i=1}^{4})/\sqrt{4}$.}
    \label{fig:S1}
\end{figure}

We numerically integrate the particle dynamics defined by Eq.~(1) of the main text using a custom \texttt{Julia} code based on a finite-difference time discretization (Euler scheme). Particles evolve in a two-dimensional square domain of linear size $L = 200\,\kappa_0$, with periodic boundary conditions applied along both spatial directions. At the initial time, particle positions are uniformly distributed within the domain $[0,L]\times[0,L]$, and their phases are all set to 0. We vary the density $\rho$ keeping fixed the other parameters: $\varepsilon = 15$, $\mu_r = 1$, $\mu_\phi = 1$, $D_r = 1$, $D_\phi = 1$, $\omega = 10$, $\sigma_0 = 0.5$, $U_0 = 1$, and $\lambda = 0.1$ with timestep $dt = 1\times 10^{-3}$. The quantity $\Upsilon$ defined as
\begin{equation}
\label{eq:Upsilon}
\Upsilon = \frac{1}{N \omega} \sum_{j=1}^N \langle \dot\phi_j\rangle \ ,
\end{equation}
measures the mean pulsation rate of the particles normalized by the bare driving frequency $\omega$: it equals unity when all particles cycle freely ($\langle\dot\phi_j\rangle=\omega$) and vanishes in the fully arrested state ($\rho \geq \rho_c$) where phase dynamics are frozen by repulsive interactions [Figs.~\ref{fig:S1}].

\subsection{Defect analysis}

Topological defects are detected from the particle phases using a Delaunay triangulation of particle positions. For each triangle, we compute the discrete phase residue
\begin{equation}
\phi_r = -\sum_{\langle ij \rangle} \mathrm{atan2}\bigl(\sin(\phi_i - \phi_j), \cos(\phi_i - \phi_j)\bigr),
\end{equation}
where the sum runs over the three edges of the triangle. A triangle with $|\phi_r|$ above a small threshold hosts a
topological defect on its centroid; the sign of $\phi_r$ identifies the charge as $q = +1$ or $q = -1$. Periodic duplicate detections and numerically spurious charge-imbalanced pairs are subsequently removed.

To characterize the internal structure of each defect, we collect the phase values $\phi \in [0, 2\pi)$ of all particles in the first and second neighbor shells of the defect core in many snapshots. Since the phase winds exactly once (by $2\pi$) around a defect, their ordering reflects the angular structure of the winding. We accumulate the empirical distribution $P(\phi)$ over all detected defects of the same charge sign, and compute its cumulative distribution function
\begin{equation}
F(\phi) = \int_0^{\phi} P(\phi')\mathrm{d}\phi' ,
\end{equation}
which maps phase values to angular positions $\theta = 2\pi F(\phi) \in [0, 2\pi)$ around the core. Inverting this relation yields the phase profile $\phi(\theta)$.
For a perfectly isotropic defect, $P(\phi)$ is uniform and $\phi(\theta) = \theta$ is a straight diagonal; deviations reveal an anisotropy in the defect structure [Eq. (2)].


\subsection{Defect tracking}

Defects are tracked across successive snapshots by solving a global assignment problem at each frame. A cost matrix is built between currently detected defects and active trajectories, with infinite cost assigned to opposite-charge pairs; for same-charge pairs, the cost is a soft function of the periodic distance to the trajectory's predicted position, with a steep penalty beyond a threshold $d_\mathrm{th}$. The optimal one-to-one assignment is found via the Hungarian algorithm~\cite{munkres1957}. Unmatched detections are assigned new trajectory IDs, while unmatched active trajectories are suspended for up to a configurable number of frames before being closed. After tracking, creation and annihilation events are identified by a second bipartite-matching pass: opposite-charge trajectory pairs that begin (or end) in the same frame within distance
$d_\mathrm{th}$ are labeled as nucleation (or annihilation) partners. The cost is linear in inter-defect distance for $d < d_{\rm th}$ and grows steeply (a factor 10 increase in slope) beyond $d_{\rm th}$. The distance threshold is $d_{\rm th} = 10.0\,\kappa_0$.


\subsection{Hydrodynamic simulations}
\label{subsec:hydro}

We simulate the hydrodynamics given in Eq.~(8) of the main text using a custom \texttt{Julia} code. We split each timestep $\mathrm{d}t$ into two sub-steps  using the operator splitting scheme described below.

\begin{itemize}
    \item Step 1: Compute the complex fields $f_1$ and $f_2$ from the phase $\phi$ and dispersion $\sigma$ fields as
    \begin{equation}
        f_1 = \rho \, \mathrm{sinc}(\kappa/2) \, e^{i\psi}, \quad f_2(\psi, \kappa) = \rho \, \mathrm{sinc}(\kappa) \, e^{2i\psi}.
    \end{equation}

    \item Step 2: Update $f_1$ on each lattice site using Euler time-stepping:
    \begin{equation}
    \begin{aligned}
        B({\bf r}, t+dt) = f_1(t) &+ \mathrm{d}t \left\{ \left(-D_\phi + i\omega\right) f_1 + i h(\rho) \left(f_2 - \rho\right) \right. \\
        &\left. + \frac{\varepsilon}{2}\left[\rho_0 f_1 - f_1 \,\mathrm{Re}(f_2) - i f_1 \,\mathrm{Im}(f_2)\right] \right\} + \frac{\sqrt{2 D_\phi \,\mathrm{d}t}}{\mathrm{d}x} \, \Lambda_1,
    \end{aligned}
    \end{equation}
    where $\Lambda_\phi = u_1 + i v_1$ is the complex noise term with $u_1$ and $v_1$ generated via Cholesky decomposition as described below.

    \item Step 3: Apply a semi-implicit spectral step for diffusion. In Fourier space, the field $\tilde{f}_{1,q}$ is updated as
    \begin{equation}
        \tilde{f}_{1}(q,t+\mathrm{d}t) = \frac{1 - D_r q^2 \,\mathrm{d}t}{1 + D_r q^2 \,\mathrm{d}t} \, \tilde{B}({\bf q},t).
    \end{equation}

    \item Step 4: Extract the updated phase and dispersion fields:
    \begin{equation}
        \psi = \arg(f_1), \quad \kappa = 2\,\mathrm{sinc}^{-1}\left(\frac{|f_1|}{\rho_0}\right),
    \end{equation}
    where $\mathrm{sinc}^{-1}$ is computed numerically via a precomputed lookup table using Newton-Raphson inversion. If $|f_1| > \rho_0$, we set $|f_1| = \rho_0$ to ensure $\sigma \geq 0$.

    \item Step 5: Return to Step 1.
\end{itemize}

We use a grid of size $256 \times 256$ with periodic boundary conditions. The parameters used in the simulations are $\rho_0 = 0.9075$, $D_r = 1$, $D_\phi = 0.2$, $\varepsilon = 6$, $\omega = 1$, $c_\phi = 2\rho_0^6$, and $\mathrm{d}t = 10^{-3}$. The system size is $L = 150$ with $N = 256$ grid points and $dx = L/N$. To generate the two random variables $u_1$ and $v_1$ with the covariance matrix given in Eq.~(13) of the main text, we employ the Cholesky factorization method described in Ref.~\cite{Glasserman}. Given a positive definite covariance matrix $\Sigma$, we seek a lower triangular matrix $A$ such that $AA^T = \Sigma$. If $Z_1, Z_2$ are independent standard normal random variables, then the transformation $(u_1, v_1)^T = A(Z_1, Z_2)^T$ yields random variables with the desired covariance structure.
This yields
\begin{align}
    u_1 &= \frac{1}{\sqrt{2}}\sqrt{f_0 - \mathrm{Re}[f_{2}]} \, Z_1, \\
    v_1 &= -\frac{1}{\sqrt{2}}\frac{\mathrm{Im}[f_{2}]}{\sqrt{f_0 - \mathrm{Re}[f_{2}]}} Z_1 + \frac{1}{\sqrt{2}}\sqrt{f_0 + \mathrm{Re}[f_{2}] - \frac{(\mathrm{Im}[f_{2}])^2}{f_0 - \mathrm{Re}[f_{2}]}} \, Z_2,
\end{align}
where $Z_1, Z_2 \sim \mathcal{N}(0,1)$ are independent standard normal random variables.

\section{HYDRODYNAMICS}
\label{sec:hydro}

\subsection{Coarse-graining procedure}
\label{subsec:coarse}
Following a  similar procedure to \cite{zhang2023}, we simplify Eq. (1) into the effective dynamics:
\begin{equation}
    \dot{\mathbf{r}}_i=\sqrt{2 D} \boldsymbol{\xi}_i ,
    \quad
    \dot{\phi}_i=\omega - h(\rho) \cos\phi_i + \sum_{j}\varepsilon \sin \left(\phi_j-\phi_i\right)\delta\left(\mathbf{r}_i-\mathbf{r}_j\right) +\sqrt{2 D_\phi} \eta_i .
\end{equation}
We then consider the global distribution $f(\mathbf{r}, \phi, t) = \sum_i \delta({\bf r}-{\bf r}_i(t))\delta(\phi-\phi_i(t))$, and by using Ito's lemma we write its time derivative as
\begin{equation}
    \partial_t f = \sum_i \left(\dot {\bf r}_i \partial_{{\bf r}_i}+\dot \phi_i \partial_{\phi_i} + D_r \partial_{{\bf r}_i {\bf r}_i}^2 +D_\phi \partial_{\phi_i\phi_i}^2 \right)f_i ,
    \quad
    f_i = \delta({\bf r}-{\bf r}_i)\delta(\phi-\phi_i) ,
\end{equation}
which leads to
\begin{equation}\label{eq:dens}
\begin{aligned}
    \partial_t f= & -\partial_\phi\Big\{ f(\mathbf{r}, \phi, t) \Big( \omega+\sum_j\Big[\varepsilon\sin \left(\phi_j-\phi\right) - h(\rho) \cos \phi\Big] \delta\left(\mathbf{r}-\mathbf{r}_j\right)\Big)\Big\}
    \\
    & -\sum_i\left[\sqrt{2 D} \boldsymbol{\xi}_i \cdot \partial_{\mathbf{r}} f_i+\sqrt{2 D_\theta} \eta_i \partial_\theta f_i\right]+\left(D \partial_{\mathbf{r r}}^2 + D_\phi \partial_{\phi\phi}^2\right) f \ .
\end{aligned}
\end{equation}
Introducing the modes $f_n$ of the joint density $f$:
\begin{equation}
    f_n(\mathbf{r}, t)=\int \mathrm{d} \phi e^{\mathrm{i} n \phi} f(\mathbf{r}, \phi, t)
\end{equation}
we deduce Eq.~(4) of the main text by multiplying Eq.~\eqref{eq:dens} with $e^{in\phi}$ and integrating over $\phi$, where the noise term $\Lambda_n$ reads
\begin{equation}
    \Lambda_n=\sum_j e^{\mathrm{i} n \phi_j}\left[-\sqrt{2 D} \boldsymbol{\xi}_j \cdot \partial_{\mathbf{r}} \delta\left(\mathbf{r}-\mathbf{r}_j\right)+\mathrm{i} n \sqrt{2 D_\phi} \eta_j \delta\left(\mathbf{r}-\mathbf{r}_j\right)\right] .
\end{equation}
Differently from Ref.~\cite{zhang2023} where an adiabatic closure is used, we now find a closed equation for $f_1$ by using the relation $f_2(f_1)= \rho\sinc(2\sinc^{-1}(|f_1|/\rho))e^{i\arg(f_1)}$.


\subsection{Homogeneous solutions}
\label{subsec:homogeneous}

To study the stationary homogeneous solutions of Eq.~(8) in the main text, we neglect the spatial and fluctuating terms. Then, by separating the real and imaginary parts, we obtain the dynamics of $(\psi,\kappa)$ that define $f_1$ [see Eq.~(7) in main text]
\begin{equation}
\begin{aligned}
    &\dot \psi =   \omega - h(\rho)\cos\psi \frac{\sinc(\kappa)+1}{2\sinc(\kappa/2)} \ ,
    \\
    &\dot \kappa = \frac{1}{M(\kappa)} \left[ \frac{h(\rho)}{2}\sin\psi\,(\sinc(\kappa)-1) -\dfrac{\varepsilon}{2}\rho \, \sinc\left(\dfrac{\kappa}{2}\right) \left(\sinc(\kappa) - 1\right) - D_\phi \, \sinc\left(\dfrac{\kappa}{2}\right) \right] \ ,
\end{aligned}
\label{eq:phi_sigma}
\end{equation}
where

\begin{equation}
    M(\kappa) = \dfrac{1}{\kappa} \left[ \cos\dfrac{\kappa}{2}-\sinc\dfrac{\kappa}{2} \right] \ .
\end{equation}
When $D_\phi = 0$ and $\omega \leq h(\rho)$, there exist two fixed points $(\bar \psi, \bar\kappa) =  (\pm\arccos{\left(\omega/  h(\rho)\right)}, 0)$. The stable fixed point is determined by linear stability analysis. $(\psi^*, \kappa^*) = (-\arccos\left(\omega/h(\rho)\right), 0)$ is stable since $\left.\partial_\psi \dot{\psi}\right|_{\psi^*} = h(\rho)\sin\psi^* < 0$. On the other hand, when $\omega > \rho$ there exist one cycling solution with constant $\kappa = 0$. In this case, Eq.~\eqref{eq:phi_sigma} reduces to the ODE
\begin{equation}
    \dot \psi = \omega -  h(\rho) \cos \psi ,
\end{equation}
which we can solve explicitly as
\begin{equation}
   \int_{\psi(0)}^{\psi(t)}\dfrac{1}{\omega -h(\rho) \cos(\psi)}d\psi = \int_0^td\tau ,
\end{equation}
yielding
\begin{equation}
    \psi(t) = 2 \tan^{-1} \left[ \frac{\sqrt{\omega^2 -h(\rho)^2}}{\omega+h(\rho)} \tan \left( \frac{\sqrt{\omega^2 -h(\rho)^2}}2 t \right) \right] ,
\end{equation}
where we assumed $\psi(0) = 0$. We define the critical density $\rho_c$ as the value of $\rho$ that satisfies $\omega = h(\rho_c)$.

\begin{figure}
    \centering
    \includegraphics[width=0.5\columnwidth]{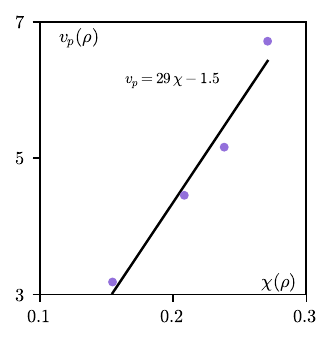}
    \caption{Parametric plot of defect peak velocity $v_{p}(\rho)$ and defect asymmetry $\chi(\rho)$ as functions of density $\rho$. The data is deduced from Fig.~1(h) of the main text, and the linear fit (solid black line) corresponds to Eq.~\eqref{eq:vpfit}.}
    \label{fig:S2}
\end{figure}


\subsection{Effective defect dynamics}

Equation~(13) of the main text describes each defect as a chiral active particle with density-dependent motility $\nu(\rho)$, angular velocity $\Omega(\rho)$, translational diffusivity $D_t = 1$, and inverse persistence $1/\tau= 0.075$. The angular velocity $\Omega(\rho) = \sqrt{\omega^2 - h(\rho)^2}$ is taken directly from Eq.~(12) of the main text. To estimate motility $\nu(\rho)$, we use the procedure described below.

For each particle density $\rho_{\rm part}$, we compute the reduced distance to the arrest transition $\delta\rho = (\rho_c^{\rm part} - \rho_{\rm part})/\rho_c^{\rm part}$, using the critical density $\rho_c^{\rm part}$ of the particle-based model (Sec.~S1A). The corresponding hydrodynamic density $\rho_{\rm hydro}$ is obtained by inverting
\begin{equation}
\delta\rho = \frac{\rho_c^{\rm hydro} - \rho_{\rm hydro}}{\rho_c^{\rm hydro}} ,
\end{equation}
using the critical density $\rho_c^{\rm hydro}$ of the hydrodynamic model (Sec.~S2B), so that the particle and hydrodynamic densities are matched by the same reduced distance to arrest.

In the microscopic dynamics, for each density $\rho \in \{1.70, 1.75, 1.80, 1.85\}$, the asymmetry $\chi(\rho)$ is extracted from the phase profiles of defects (Sec.~S1B), and so is the defect peak speed $v_p(\rho)$. We fit the four data pairs $(\chi,v_p)$ shown in Fig.~\ref{fig:S2} with the linear relation
\begin{equation}\label{eq:vpfit}
    v_p(\chi) = a\chi + b
\end{equation}
using ordinary least squares, yielding $a \simeq 29$ and $b \simeq -1.5$, over the range
$\chi \in [0.155, 0.271]$ (corresponding to $\rho \in [1.70, 1.85]$,
with $\rho_c = 1.86$).

In the hydrodynamics, the analytical profile $\psi(\theta)$ [Eq.~(11)] yields the asymmetry
$\chi(\rho)$, and the motility follows from the linear fit of Eq.~\eqref{eq:vpfit} as $\nu(\rho) = v_p\!\left(\chi(\rho)\right)$. Because $\rho_c^{\rm hydro} \neq \rho_c^{\rm part}$, matching by
$\delta\rho$ does not preserve $\chi$: the four values of $\rho_{\rm hydro}$ obtained this way yield systematically larger asymmetry $\chi(\rho_{\rm hydro})$ than $\chi(\rho_{\rm part})$, and correspondingly larger motility $\nu(\rho)$, than would be obtained by matching $\chi$ directly.

The mean-squared displacement $\langle \Delta x^2 \rangle(t)$ shown in Fig.~3(e) of the main text is obtained by direct Euler--Maruyama integration of Eq.~(13) with time step $dt = 0.03$ and recording interval $dt_{\rm rec} = 0.15$ (5 integration steps per stored frame). For each of the four $(\nu, \Omega)$ pairs, $N_{\rm traj} = 300$ independent trajectories are propagated for $n_{\rm stored} = 333$ frames; the mean-squared displacement at lag $d$ is estimated as the mean of $|\mathbf{r}(t+d) - \mathbf{r}(t)|^2$ pooled over all (trajectory, time-origin) pairs, with standard error ${\rm SE} = \sigma/\sqrt{N}$, where $N$ is the number of pairs and $\sigma$ their standard deviation.

\end{document}